%% file: main.tex
\documentclass[sigconf]{acmart}

\newcommand{\niparagraph}[1]{\noindent\textbf{#1}}

\newcommand{\sysname}[0]{\textsc{Pimba}}

\definecolor{customblue}{rgb}{0.20, 0.35, 0.91}
\newcommand{\bluetext}[1]{\textcolor{customblue}{#1}}

\usepackage{acmart-taps}
\usepackage{enumitem}
\usepackage{multirow}
\usepackage{verbatim}

\AtBeginDocument{%
  }

\copyrightyear{2025}
\acmYear{2025}
\setcopyright{cc}
\setcctype{by}
\acmConference[MICRO '25]{58th IEEE/ACM International Symposium on
Microarchitecture}{October 18--22, 2025}{Seoul, Republic of Korea}
\acmBooktitle{58th IEEE/ACM International Symposium on
  Microarchitecture (MICRO '25), October 18--22, 2025, Seoul, Republic
of Korea}\acmDOI{10.1145/3725843.3756121}
\acmISBN{979-8-4007-1573-0/2025/10}

\begin{document}

\title{\sysname{}: A Processing-in-Memory Acceleration for
Post-Transformer Large Language Model Serving}

\author{Wonung Kim}
\affiliation{
  \institution{KAIST}
  \city{Daejeon}
  \country{Republic of Korea}
}
\email{wukim@casys.kaist.ac.kr}

\author{Yubin Lee}
\affiliation{
  \institution{KAIST}
  \city{Daejeon}
  \country{Republic of Korea}
}
\email{yblee@casys.kaist.ac.kr}

\author{Yoonsung Kim}
\affiliation{
  \institution{KAIST}
  \city{Daejeon}
  \country{Republic of Korea}
}
\email{yskim@casys.kaist.ac.kr}

\author{Jinwoo Hwang}
\affiliation{
  \institution{KAIST}
  \city{Daejeon}
  \country{Republic of Korea}
}
\email{jwhwang@casys.kaist.ac.kr}

\author{Seongryong Oh}
\affiliation{
  \institution{KAIST}
  \city{Daejeon}
  \country{Republic of Korea}
}
\email{sroh@casys.kaist.ac.kr}

\author{Jiyong Jung}
\affiliation{
  \institution{KAIST}
  \city{Daejeon}
  \country{Republic of Korea}
}
\email{jyjung@casys.kaist.ac.kr}

\author{Aziz Huseynov}
\affiliation{
  \institution{KAIST}
  \city{Daejeon}
  \country{Republic of Korea}
}
\email{aziz@casys.kaist.ac.kr}

\author{Woong Gyu Park}
\affiliation{
  \institution{KAIST}
  \city{Daejeon}
  \country{Republic of Korea}
}
\email{wgpark@casys.kaist.ac.kr}

\author{Chang Hyun Park}
\affiliation{
  \institution{Uppsala University}
  \city{Uppsala}
  \country{Sweden}
}
\email{chang.hyun.park@it.uu.se}

\author{Divya Mahajan}
\affiliation{
  \institution{Georgia Institute of Technology}
  \city{Atlanta}
  \state{GA}
  \country{USA}
}
\email{divya.mahajan@gatech.edu}

\author{Jongse Park}
\affiliation{
  \institution{KAIST}
  \city{Daejeon}
  \country{Republic of Korea}
}
\email{jspark@casys.kaist.ac.kr}

\renewcommand{\shortauthors}{Kim et al.}

\renewcommand{\arraystretch}{0.85}


\include{body/abstract}



\maketitle

\input{body/introduction}

\input{body/background}

\input{body/characterization}

\input{body/principle}

\input{body/architecture}

\input{body/evaluation}

\input{body/related_works}

\input{body/discussion}

\input{body/conclusion}

\input{body/ack}



\balance{}
\bibliographystyle{ACM-Reference-Format}
\bibliography{main}

\input{body/ae}
\end{document}

%% file: body/abstract.tex
\begin{abstract}
  Transformers are the driving force behind today's Large Language
  Models (LLMs), serving as the foundation for their performance and
  versatility.
  Yet, their compute and memory costs grow with sequence length,
  posing scalability challenges for long-context inferencing.
  In response, the algorithm community is exploring alternative
  architectures—such as state space models (SSMs) (e.g., Mamba-2),
  linear attention, and recurrent neural networks (RNNs)—which we
  refer to as \emph{post-transformers}.
  This shift presents a key challenge: building a serving system that
  efficiently supports not only emerging post-transformer LLMs but
  also existing transformer models within a unified framework.
  To address this challenge, we analyze the performance
  characteristics of transformer and post-transformer LLMs.
  Despite their algorithmic differences, both are largely
  bounded by memory bandwidth under batched inference—due to
  attention in transformers and state updates in post-transformers.
  Inspired by this finding, we propose \sysname{}, an accelerator
  solution that aims to address the memory bottleneck by jointly
  leveraging (1) Processing-in-Memory (PIM) paradigm and (2) LLM quantization.
  Further analyses suggest two additional insights: (1) state update
  operations, unlike attention, incur high hardware cost, making
  per-bank PIM acceleration inefficient, and (2) different
  low-precision arithmetic methods offer varying accuracy-area
  tradeoffs, while we identify Microsoft's MX as a Pareto-optimal choice.
  Building on these insights, we design the architecture of
  \sysname{} as an array of
  State-update Processing Units (SPUs), each shared between two banks
  to enable interleaved access.
  Each SPU includes a State-update Processing Engine (SPE) that
  comprises element-wise multipliers and adders using MX-based
  quantized arithmetic, enabling efficient execution of state update
  and attention operations.
  Our evaluation shows that, compared to LLM-optimized GPU and
  GPU+PIM systems, \sysname{} achieves up to 4.1$\times$ and
  2.1$\times$ higher generation throughput, respectively.

\end{abstract}

%% file: body/introduction.tex
\section{Introduction}
Every industrial and enterprise sector in our society is either
actively using Large Language Models (LLMs) or eager to adopt
them~\cite{intro_1, intro_2}.
LLM's widespread success can be attributed to the effectiveness of
their core algorithmic component,
\emph{transformers}~\cite{vaswani2023attention}.
While transformers offer remarkably versatile capabilities and continue to
dominate LLMs, their enormous resource demands are a significant
concern for LLM providers.
Transformer-based LLMs scale quadratically in compute and linearly in
memory footprint with sequence length, while emerging
applications--such as test time scaling~\cite{deepseekr1, googletts,
o1, s1}, retrieval-augmented generation~\cite{rag, pmlrrag, copy,
toolformer}, and multimodal input fusion~\cite{gpt4o, llama4}--are
driving demand for longer sequence lengths, recently reaching up to 2
billion tokens in industry-leading models~\cite{longrope}.
Moreover, batched inferencing exacerbates these resource demands,
forcing hyperscalers to invest billions of dollars in equipping their
data centers with hundreds of thousands of costly GPUs, each priced
at \$30,000 or more~\cite{meta-h100-purchase}.

\begin{figure}[t]
  \centering
  \includegraphics[width=0.95\linewidth]{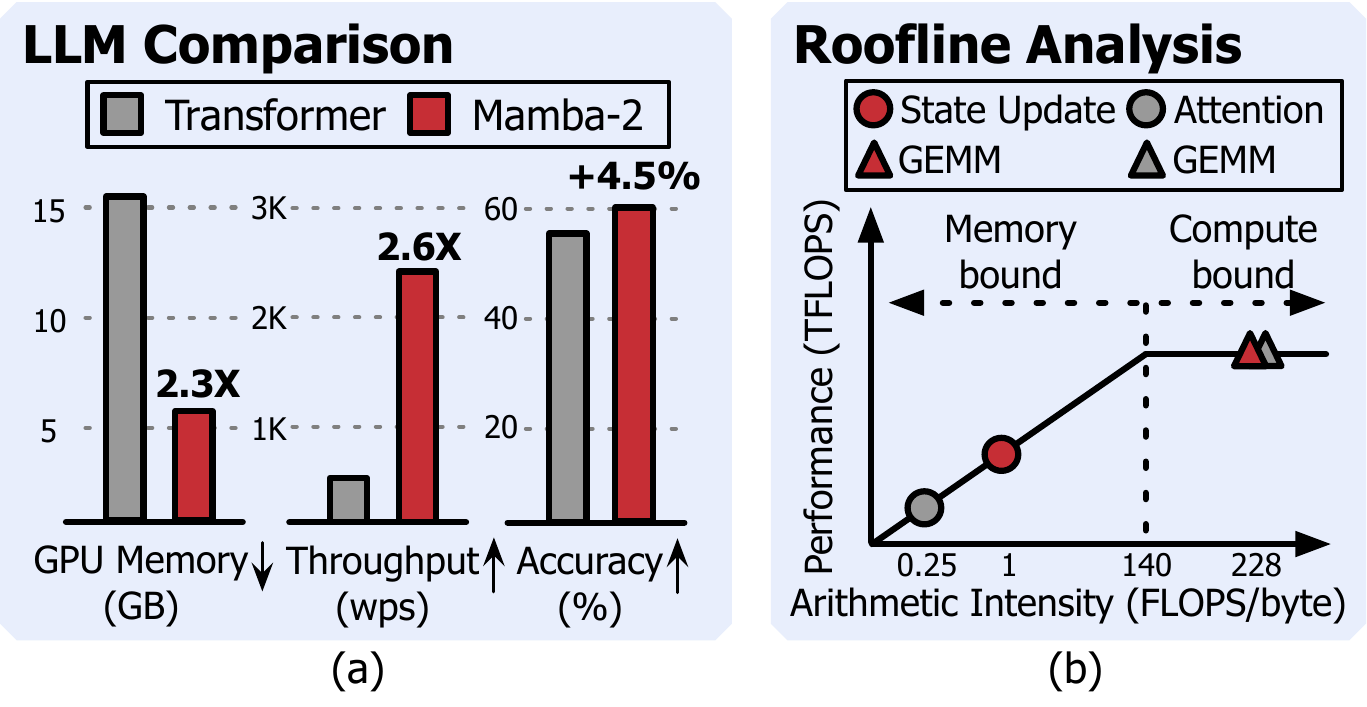}
  \caption{Comparison between 2.7B parameter transformer and Mamba-2.
  Accuracy results are referenced from~\cite{mamba2}.}
  \label{fig:mamba_vs_transformer}
\end{figure}

Recently, the algorithm community has actively explored alternative
approaches, including state space models (SSMs), linear attention
mechanisms, and recurrent neural networks (RNNs).
In this paper, we henceforth refer to LLMs employing these
alternative algorithmic techniques as ``post-transformer'' LLMs.
We argue that post-transformer LLMs have the potential to serve as a
promising complement to transformer-based LLMs, providing comparable
algorithmic capabilities with significantly lower and constant
resource demands~\cite{mamba, mamba2, s4, h3, hyena, s4d, hippo,
  lssl, linear_attention, gated_linear_attention, gsa, retnet, rwkv,
rwkv6, xlstm, hgrn2}.
Figure~\ref{fig:mamba_vs_transformer}(a) presents an empirical
evidence that supports our argument.
The figure compares the memory usage, throughput, and accuracy of a
transformer-based LLM with a post-transformer LLM, Mamba-2, both
having a model size of 2.7 billion parameters\footnote{Detailed
experimental methodology is presented in Section~\ref{sec:method}}.
The results show that Mamba-2 requires 2.3$\times$ less memory
capacity, delivering 2.6$\times$ higher throughput than the
transformer counterpart, while achieving 4.5\% higher accuracy.
Despite this considerable potential, the architecture and system
community have a limited understanding of the implications of these
algorithms, causing LLM providers to hesitate in adopting
them for their serving systems.
To this end, this paper sets out to bridge this gap through a
comprehensive workload analysis and performance characterization, and
to devise a solution that leverages the resulting insights.
The first of these insights is that many post-transformer LLMs share a
common algorithmic operation, \emph{state update}, which propagates
and evolves contextual information across tokens.
This commonality offers a promising opportunity for architectural
generalization and acceleration.
We also discovered that similar to the attention operation in
transformer-based LLMs, this state update operation becomes the
performance bottleneck due to its low arithmetic intensity.
Figure~\ref{fig:mamba_vs_transformer}(b) reports the roofline
analysis results that the arithmetic intensity of state update
operation is 4$\times$ larger than that of attention, while it is
still significantly bandwidth-bound.
Inspired by these insights, we propose \sysname{}, an acceleration
solution that addresses the memory bandwidth bottleneck by jointly
exploiting (1) Processing-in-Memory (PIM) paradigm, and (2) LLM quantization.
While prior works have extensively investigated these techniques for
transformer-based LLM serving~\cite{xiao2023smoothquant,
  hooper2024kvquant, monkey, atom, heo2024neupims,
park2024attacc,ianus}, we observe that post-transformer LLMs
demonstrate significantly different behaviors, requiring distinct
design choices to enable a unified serving system that accommodates
the two classes of LLM architectures.
Below, we share the empirical insights and their corresponding
principles that govern our accelerator design:
\begin{itemize}[labelindent=0.5em,nolistsep,leftmargin=1.0em]
  \item \textbf{(Principle 1): Maximizing hardware resource sharing
    for area efficiency:}
    Existing LLM-targeted PIM acceleration methods~\cite{hbm_pim,
    newton, heo2024neupims, park2024attacc, ianus} focus on
    supporting matrix-vector multiplication (i.e., GEMV) since
    attention operation consists of a full of GEMVs.
    However, this approach is unsuitable for post-transformer
    algorithms since implementing the state update operation in
    hardware incurs significantly larger area costs due to the
    variety of primitives in state update operation, such as
    element-wise multiplication, element-wise addition, and vector dot products.
    Thus, in designing \sysname{}, we aim to exploit the hardware
    resource sharing for maximizing area efficiency.
  \item \textbf{(Principle 2): Achieving both accuracy and
    area-efficiency from low-precision arithmetic:}
    While quantizing the \emph{state} in post-transformers can reduce
    computation cost and memory footprint, it also affects area efficiency.
    We also discover that, due to the state ``update’’ mechanism,
    conventional numerical formats cause severe accuracy degradation,
    rendering them impractical for post-transformers.
    We carefully explore the accuracy-area tradeoffs and observe that
    different low-precision arithmetic approaches exhibit different
    characteristics.
    We thoroughly perform an empirical study to understand the
    differences and aim to employ a Pareto-optimal quantization
    technique for our solution.
\end{itemize}
\noindent Building upon these two principles, we design the \sysname{}
accelerator architecture, which incorporates the following key elements:
\niparagraph{State-update Processing Unit (SPU).}
At the core of \sysname{} is the State-update Processing Unit (SPU),
which includes a State-update Processing Element (SPE).
Deploying an SPE for each bank would incur excessive area costs and
reduce memory capacity, rendering this approach impractical under the
stringent area constraints of PIM compute units.
To address this, \sysname{} assigns one SPU to every two banks.
The SPU alternates between reading from and writing to the row
buffers of different banks, performing computations in an interleaved manner.
This design sustains throughput while optimizing area efficiency.
\niparagraph{SPE with MX-based quantized arithmetic.}
Empirical analysis suggests that among various quantization formats,
MX8~\cite{rouhani2023shared} (requiring an average of 8 bits per
value) emerges as a Pareto-optimal choice in the accuracy-efficiency
tradeoff, while aligning seamlessly with memory address alignment requirements.
This enables area- and power-efficient implementation of SPEs within
the constraints of PIM.
Consequently, we design custom MX8 vector multipliers and adders,
significantly improving resource efficiency.
\niparagraph{End-to-end \sysname{} system design.}
We construct the \sysname{} system by jointly leveraging \sysname{}
accelerators with GPUs, offloading state update and attention
operations to PIM, while delegating other tasks to GPUs.
\sysname{} includes custom DRAM commands and command scheduling
techniques to manage state pre-charging and subsequent generative computations.
Our PIM accelerator and its interface use a system architecture
similar to the existing PIM-based LLM serving
systems~\cite{heo2024neupims, park2024attacc, ianus, cxl-pnm},
allowing \sysname{} to serve as a ``drop-in replacement'' in
transformer-serving systems adapted to support post-transformer LLMs as well.
To evaluate \sysname{}'s effectiveness, we use four post-transformer
LLM models--Mamba-2, GLA, RetNet, and HGRN2~\cite{mamba2,
gated_linear_attention, retnet, hgrn2}-- along with Zamba2, a hybrid
transformer-Mamba-2 model~\cite{zamba2}, and OPT~\cite{zhang2022opt},
a traditional attention-based model.
Our experimental results show that compared to LLM-optimized GPU and
GPU+PIM systems, \sysname{} achieves 14.6$\times$ and 6.9$\times$
lower latency in state update operations, resulting in up to
4.1$\times$ and 2.1$\times$ higher throughput, respectively, with
minimal area overhead on the memory device.
These advantages in both performance and area-efficiency demonstrate
that \sysname{} is an effective PIM-based solution for LLM serving,
capable of supporting both transformer and post-transformer models,
paving the way toward scalable and cost-efficient deployment of
emerging LLM architectures.
A full-system simulator for \sysname{} and the accuracy evaluation code
are open-sourced at \bluetext{\url{https://github.com/casys-kaist/pimba}}.
%

%% file: body/background.tex
\section{Background}
\label{sec:background}
\subsection{Transformer-based LLMs and their Limitations}
\label{subsec:background_llm}
\begin{figure}[t]
  \centering
  \includegraphics[width=0.8\linewidth]{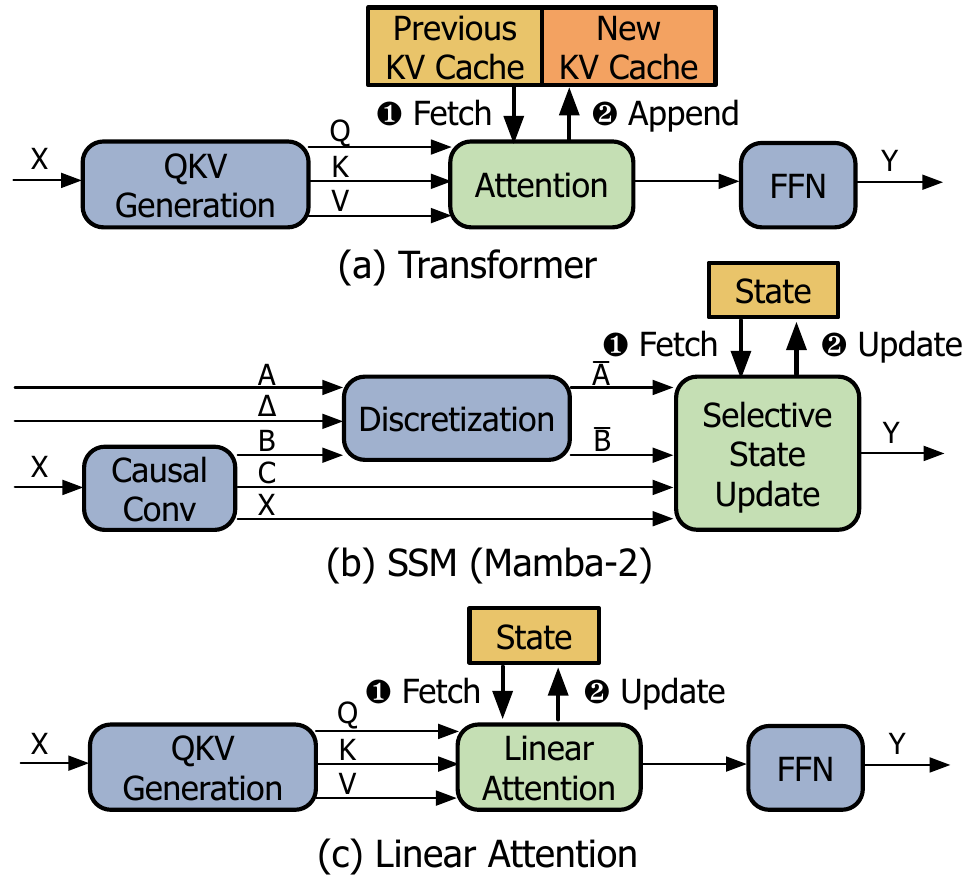}
  \caption{Model architectures of (a) Transformer, (b) Mamba-2 of
    SSM, and (c) Linear Attention. For simplicity, we focus on the
    key operations.
  }
  \label{fig:LLM}
\end{figure}
\niparagraph{Transformer-based LLMs.}
Transformers offer remarkable performance due to their attention
mechanism, which enables efficient modeling of inter-token
dependencies~\cite{llama2, devlin2019bert, Rub1, Rub2}.
Figure~\ref{fig:LLM}(a) shows the model architecture for transformer-based LLMs.
In the attention mechanism, each token in the input sequence is
projected into three distinct vectors: query ($Q$), key ($K$), and value ($V$).
The query and key vectors are used to compute the attention scores by
taking the scaled dot product, and the value vectors are used to
perform a weighted sum over these scores.
\niparagraph{Limitations.}
The auto-regressive nature of LLM requires revisiting all previous
tokens, resulting in redundant computation.
Key-Value (KV) cache is employed to prevent recomputing previous
tokens, but the transformers face the following limitations:
\aptLtoX{\begin{itemize}
  \item[(1)] \textbf{Memory usage.}
     The KV cache grows linearly with the sequence length. Despite
    prior work on enhancing memory
    efficiency~\cite{hooper2024kvquant, yang2024token, dong2024qaq,
    kang2024gear,vllm}, the fundamental property of the algorithm is
    unchanged. This ends up consuming significant amounts of GPU
    memory and imposing limits on the sequence length or the batch size.
    \item[(2)] \textbf{Latency.}
    The computation of attention layers increases linearly, even with
    the KV cache, leading to increased latency with longer sequences.
    In multi-user serving scenarios, typically processed in batches,
    the difference in compute latency can hinder efficient
    scheduling~\cite{orca,vllm}.
\end{itemize}}{\begin{description}[labelindent=0.0em,nolistsep,leftmargin=1.5em]
  \item[(1)] \textbf{Memory usage.}
    The KV cache grows linearly with the sequence length. Despite
    prior work on enhancing memory
    efficiency~\cite{hooper2024kvquant, yang2024token, dong2024qaq,
    kang2024gear,vllm}, the fundamental property of the algorithm is
    unchanged. This ends up consuming significant amounts of GPU
    memory and imposing limits on the sequence length or the batch size.

  \item[(2)] \textbf{Latency.}
    The computation of attention layers increases linearly, even with
    the KV cache, leading to increased latency with longer sequences.
    In multi-user serving scenarios, typically processed in batches,
    the difference in compute latency can hinder efficient
    scheduling~\cite{orca,vllm}.
\end{description}}

\subsection{Post-Transformer LLMs}
\label{subsec:ssm_based_llms}
Recently, alternative architectures including state space models
(SSMs)~\cite{mamba, s4, h3, hyena, s4d, hippo, lssl, mamba2}, linear
attention mechanisms~\cite{retnet, linear_attention,
gated_linear_attention, gsa}, and recurrent neural networks
(RNNs)~\cite{hgrn2, rwkv, rwkv6, xlstm} have emerged as promising
substitutes for transformers.
These \emph{post-transformer} models offer comparable capabilities to
transformers, while requiring constant resources regardless of
sequence length, addressing the fundamental limitations of transformers.
\niparagraph{SSM.}
Among the alternatives, state space models (SSMs) have demonstrated
their effectiveness, leveraging structured state transitions to
efficiently capture long-range dependencies.
The state-of-the-art, Mamba-2~\cite{mamba2}, achieves leading
performance among SSMs through its selection mechanism that
efficiently propagates prior token information.
Given the prominence of Mamba-2 in language modeling and its adoption
in numerous new models~\cite{huang2024mlmamba,
nvidia-mamba2, zamba2, codestral}, this
paper focuses on Mamba-2 as a representative of SSMs.
Figure~\ref{fig:LLM}(b) illustrates the key operations of Mamba-2.
Among these, the selective state update operation is the core in
Mamba-2, which operates with $H$ parallel heads, akin to the
multi-head attention mechanism in transformers.
The inputs to the selective state update include vectors
$\overline{A}$, $\overline{B}$, $C$, and $X$.
These are partitioned across the $H$ heads, yielding scalar $a^h$ and
vectors $\overline{B^h}$, $C^h$, and $X^h$ for each head.
Each head maintains its own state matrix, which is updated through
the following steps at each time step:
\aptLtoX{\begin{itemize}
  \item[(1)] \textbf{State decay.} The previous state matrix is
    decayed by multiplying it with scalar $a^h$, decaying the
    influence of older information.
  \item[(2)] \textbf{Outer product.} The outer product of vectors
    $\overline{B^h}$ and $X^h$ is computed, capturing the
    interactions between these vectors.
  \item[(3)] \textbf{Update.} The resulting outer product matrix is
    added to the decayed state to form the updated state.
  \item[(4)] \textbf{Output.} A GEMV operation between the updated
    then transposed state matrix and vector $C^h$ produces the output vector.
\end{itemize}}{\begin{description}[labelindent=0.0em,nolistsep,leftmargin=1.5em]
  \item[(1)] \textbf{State decay.} The previous state matrix is
    decayed by multiplying it with scalar $a^h$, decaying the
    influence of older information.
  \item[(2)] \textbf{Outer product.} The outer product of vectors
    $\overline{B^h}$ and $X^h$ is computed, capturing the
    interactions between these vectors.
  \item[(3)] \textbf{Update.} The resulting outer product matrix is
    added to the decayed state to form the updated state.
  \item[(4)] \textbf{Output.} A GEMV operation between the updated
    then transposed state matrix and vector $C^h$ produces the output vector.
\end{description}}
In this sequence, each head effectively updates its internal state
based on inputs, producing outputs that contribute to the model's
overall computation.

\begin{figure}[t]
  \centering
  \includegraphics[width=\linewidth]{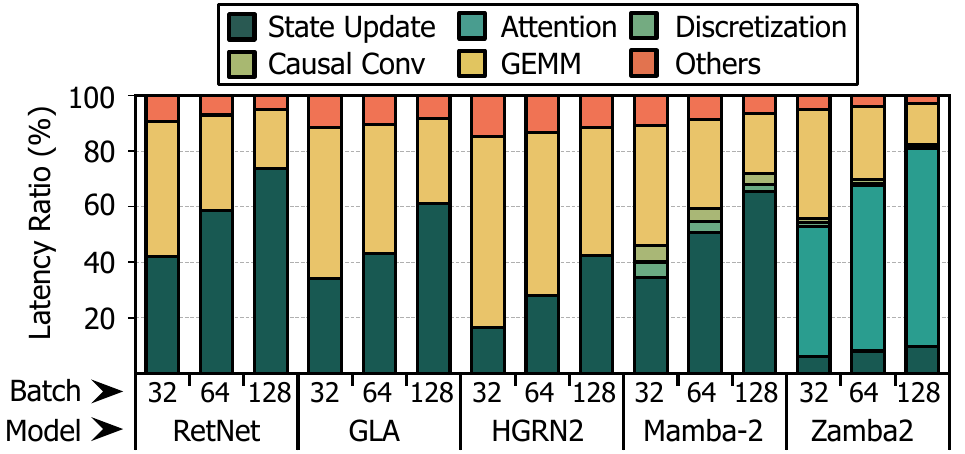}
  \caption{Latency breakdown of operations during generation phase on
    various SU-LLMs. For RetNet, GLA, HGRN2, and Mamba-2, we use a
    single generation phase due to their constant-time behavior. For
  Zamba2, we use (2,048, 2,048) input/output lengths.}
  \label{fig:motivation_breakdown}
\end{figure}

\niparagraph{Linear Attention.}
One approach is to propose entirely new architectures, such as SSM.
Alternatively, modifying the existing attention mechanism offers
another way to address the limitations of transformers.
Among such approaches, linear attention~\cite{linear_attention,
retnet, gated_linear_attention, gsa} has garnered significant
interest, as it replaces the softmax function in attention with a
linear function.
Since most linear attention mechanisms use the identity function as
the linear function, it can be expressed as
Equation~\ref{eq:linear_attention} and is illustrated in
Figure~\ref{fig:LLM}(c).
\begin{equation}
  LinearAttention(Q, K, V) = Q \cdot (K^T \cdot V)
  \label{eq:linear_attention}
\end{equation}
During the generation phase, the $K^T \cdot V$ product is used as the
state, which is continuously updated.
This is a constant-size state that does not grow with sequence length
and corresponds to steps (2)-(3) of the selective state update
operations in SSMs.
Multiplication of the state with $Q$ corresponds to the final step (4).
When a scalar decay factor is applied (e.g., RetNet~\cite{retnet}),
the linear attention mechanism aligns with the selective state update
operation in Mamba-2.
Conversely, applying an input-dependent gating mechanism (e.g.,
GLA~\cite{gated_linear_attention}) replaces the scalar decay factor
with a gating vector, which is broadcast and multiplied element-wise
with the state.
In short, both RetNet and GLA share the same or very similar state
update operation as Mamba-2.

\niparagraph{RNN.}
RNNs are being actively revisited as an alternative to transformers
for their linear computational complexity~\cite{rwkv, rwkv6, hgrn2, xlstm}.
Among these, HGRN2~\cite{hgrn2} introduces a novel architecture by
extending the conventional RNN state representation from a
one-dimensional to a two-dimensional state using an outer
product-based approach.
Interestingly, this operation closely resembles step (2) of the
selective state updates.
The forget gate in HGRN2 functions similarly to the decay mechanism,
while it employs a forget gate vector instead of a decaying scalar, akin to GLA.
\niparagraph{Combining with attention.}
Although aforementioned architectures demonstrate strong performance,
they often fall short in in-context learning, particularly in
recalling previous tokens~\cite{nvidia-mamba2}.
This limitation has motivated a body of work~\cite{nemotron-h,
zamba2, hymba, samba, nvidia-mamba2} exploring hybrid models that
combine the efficiency of alternative architectures and the
expressiveness of attention.
Notably, Nemotron-H~\cite{nemotron-h} and Zamba2~\cite{zamba2}
integrate attention layers with Mamba-2 architectures to leverage the
complementary strengths of both approaches.
By sparsely inserting attention layers, for example, one attention
layer per six Mamba-2 layers in Zamba2~\cite{zamba2}, these models
effectively restore the in-context learning capability of standard
Transformers, while maintaining the computational efficiency.

\subsection{DRAM and Processing-in-Memory (PIM)}

\niparagraph{DRAM architecture.}
DRAM is organized hierarchically, starting with channels, each
divided into ranks, which are further subdivided into bank groups.
Each bank group consists of multiple banks, with each bank storing
data in a matrix format.
Accessing data from DRAM involves three critical steps:
(1) \emph{Row Access}: the sense amplifier of the bank activates the target row.
(2) \emph{Column Access}: the specific column within the activated
row is selected, and the requested data is read out.
(3) \emph{Data Transfer}: the data is transmitted to the host via the
data bus of the DRAM channel, where only one bank of the channel can
transfer data at a time.
\niparagraph{PIM.}
Processing-in-Memory (PIM) is a realization of the Near-Data
Processing (NDP) paradigm, which has branched into various research directions.
Among these, industry-leading memory manufacturers focus on in-bank
PIM technologies, where each DRAM bank is equipped with small compute
logic to overcome the bandwidth constraints of the DRAM channel.
These accelerators perform PIM operations during the first two steps
of DRAM access, with computation handled by the in-bank logic instead
of transferring data over the bus.
As DRAM comprises multiple banks, its internal bandwidth is
significantly higher than the channel bandwidth, creating
opportunities for PIM to leverage.
Thus, PIM delivers substantial speedups for memory-bound tasks with
low arithmetic intensity.
%

%% file: body/characterization.tex
\section{Workload Characterization}
\label{sec:motivation}

\subsection{Analysis of Post-Transformer LLMs}
\label{subsec:state_update_characterization}
\niparagraph{Common operational structure.}
As discussed in Section~\ref{subsec:ssm_based_llms}, many
post-transformer models exhibit a shared structured pattern that is
increasingly evident in recent algorithms.
We find that we can unify this shared algorithmic commonality across
post-transformer models into a single, generalized operation, termed
\textbf{state update}.
For clarity, we refer to post-transformer models employing this state
update as \textbf{\underline{S}}tate
\textbf{\underline{U}}pdate-based \textbf{\underline{LLMs}}, or
\textbf{SU-LLMs} for short.
Equation~\ref{eq:state_update} represents the state update operation
for a single head.
\begin{equation}
  \label{eq:state_update}
  \begin{aligned}
    &S_t = d_t \odot S_{t-1} + k_t v_t^T \\
    &y_t = S_t^T q_t
  \end{aligned}
\end{equation}
Here, $d_t$, $q_t$, and $k_t$ are vectors with $dim_{head}$
dimension, while $v_t$ is a vector with $dim_{state}$ dimension.
The state is represented as a matrix of ($dim_{head} \times
dim_{state}$) dimensions.
First, the $d_t$ vector is broadcast to match the dimensions of the
state matrix, after which an element-wise multiplication is performed
to decay the state.
This decayed state is then updated by adding the outer product of
$k_t$ and $v_t$.
The updated state is then multiplied by $q_t$ using GEMV to produce
the output $y_t$.

\begin{figure*}[t]
  \centering
  \includegraphics[width=\linewidth]{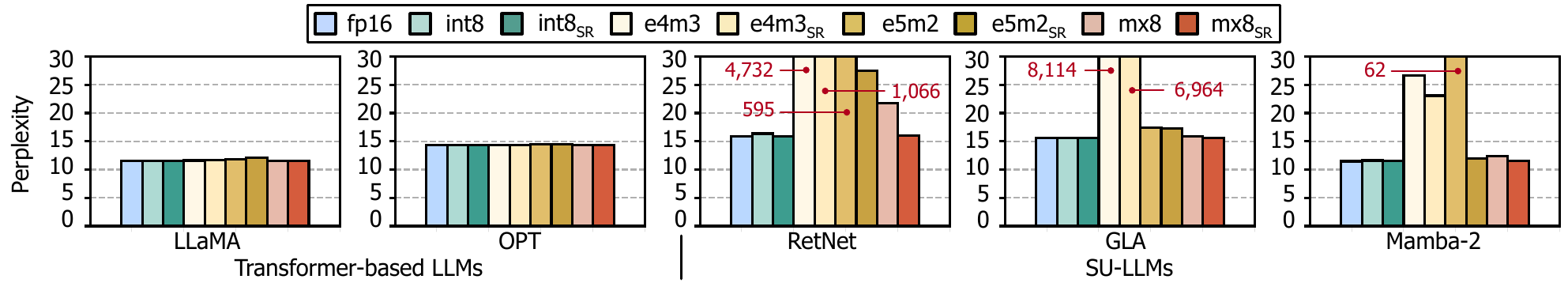}
  \caption{Perplexity of SU-LLMs and transformer-based LLMs using the
    WikTtext-2~\cite{merity2016pointer} dataset when quantized their
  respective representations to 8-bit formats.}
  \label{fig:characterization_quant}
\end{figure*}
\niparagraph{Performance Analysis.}
Figure~\ref{fig:motivation_breakdown} illustrates the latency
breakdown of operations during generation phase across 2.7B parameter
SU-LLMs--such as RetNet, GLA, HGRN2, Mamba-2~\cite{retnet,
gated_linear_attention, hgrn2, mamba2}--along with
Zamba2~\cite{zamba2}, a 7B parameter hybrid transformer-Mamba-2 model
using A100 GPU.
Unless otherwise specified, we use models with the same parameter
count throughout this paper.
The results show that state updates dominate latency, despite having
fixed memory and compute footprints.
In RetNet, as the batch size increases from 32 to 128, the time spent
on state updates rises from 41.9\% to 73.8\%, resulting in a
significant bottleneck.
This is because state updates are memory-bound and lack parameter
reuse across user requests.
They read and write the state matrix, while each operation--such as
decay, outer product, update, and GEMV--requires FLOPs proportional
to the size of the state matrix, resulting in a low operational intensity.
Furthermore, each request must independently read, update, and write
its own state.
Consequently, their latency grows linearly with batch size, rendering
state updates a performance bottleneck at large batch sizes.
Another noteworthy observation is that in a hybrid model such as
Zamba2, although the number of Mamba-2 layers greatly exceeds that of
attention layers (e.g. 6$\times$), attention still represents a
substantial fraction of the overall latency--reaching 65.5\% at a
batch size of 128.
This is because, unlike state update operations that exhibit constant
latency regardless of sequence length, attention operations scale in
latency proportionally to sequence length, making them a dominant
bottleneck in long-sequence scenarios.
Hence, to effectively accelerate hybrid models, it is critical to
optimize not only state update operations but also attention operations.

\subsection{Quantization Analysis for SU-LLMs}
\label{subsec:challenges_quantization}
As discussed in Section~\ref{subsec:state_update_characterization},
state update operations are memory-bound, leading to significant
memory bandwidth pressure.
Quantizing the state may offer a promising solution to mitigate this
issue by reducing data precision and thus memory bandwidth needs.
Although significant research has been dedicated to quantizing the KV
cache in transformer-based LLMs~\cite{xiao2023smoothquant,
hooper2024kvquant, atom}, the quantization of the state in SU-LLMs
has received little attention.
\niparagraph{Low precision formats.}
To address this gap, we explore various low-precision formats for
quantizing the state: (1) integer, (2) floating point, and (3) block
floating point formats.
For the integer format, we use an 8-bit integer with a scaling factor
across every 32 elements.
For the floating point format, we consider 8-bit variants:
\texttt{e4m3} (4 exponent bits and 3 mantissa bits) and \texttt{e5m2}
(5 exponent bits and 2 mantissa bits).
For the block floating point format, we employ
\texttt{MX}~\cite{rouhani2023shared}.
Specifically, we employ a variant of \texttt{MX}, called
\texttt{MX8}, where groups of 16 values share a common 8-bit
exponent, and pairs of values within each group share a 1-bit
microexponent to match the bit-width.
We also investigate the impact of rounding methods, particularly
stochastic rounding, which rounds numbers probabilistically based on
their distance from representable values.
\niparagraph{Implication of quantization for SU-LLMs.}
Figure~\ref{fig:characterization_quant} shows the perplexity of
various 2.7B parameter models when their respective
representations--state for SU-LLMs and KV cache for transformers--are
quantized using the Wikitext-2~\cite{merity2016pointer} dataset.
Our results reveal distinct quantization behaviors between these two
model types.
Transformer-based LLMs exhibit negligible perplexity increases across
all formats, while SU-LLMs exhibit a severe increase in perplexity
with floating point formats (e.g. 8,114 for GLA with \texttt{e4m3}).
This discrepancy arises from the SU-LLMs' continuous state ``update'' mechanism.
This makes them vulnerable to loss of small values during
accumulation due to limited mantissa precision, which is called
swamping effect~\cite{swamping, wang2018training}.
The 7-bit (\texttt{int8}) and 6-bit (\texttt{MX8}) mantissas provide
enough precision to mitigate swamping, whereas the 3-bit and 2-bit
mantissas in \texttt{e4m3} and \texttt{e5m2} render these formats
highly susceptible.
This finding aligns with conventional practices in training deep
learning models, wherein weights are stored at higher precisions to
reduce numerical errors~\cite{micikevicius2018mixed}.
Another notable observation is that stochastic rounding has a
substantial positive impact on SU-LLMs, in contrast to transformer-based LLMs.
For example, the perplexity of Mamba-2 in the \texttt{e5m2} format
drops dramatically from 62 to 11.9 when stochastic rounding is applied.
In SU-LLMs, stochastic rounding probabilistically preserves smaller
magnitude values that would otherwise be lost due to swamping,
thereby maintaining more information during state update scenario.
According to the results, employing stochastic rounding on
\texttt{int8} appears optimal for SU-LLMs.
However, this strategy might require re-evaluation in
area-constrained environments, such as PIM architectures.
We will discuss this in further detail in Section~\ref{subsec:area_accuracy}.

\begin{figure}[t]
  \centering
  \includegraphics[width=\linewidth]{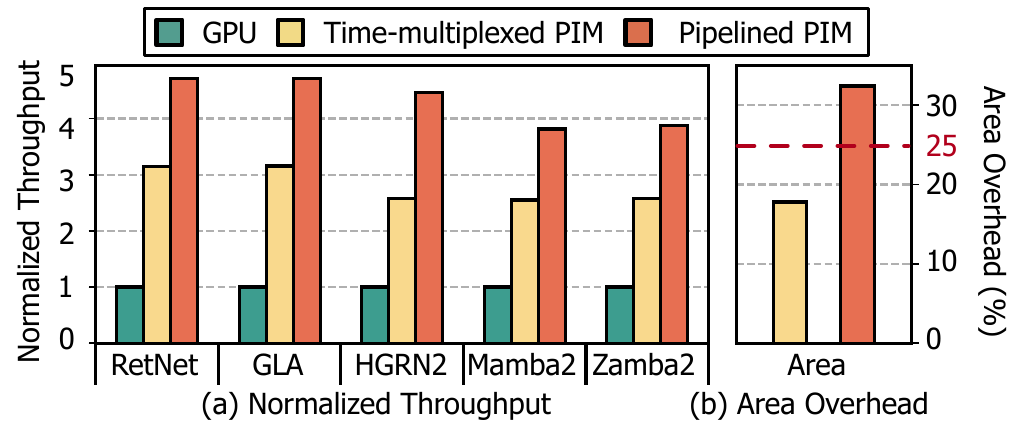}
  \caption{(a) Normalized throughput for state updates of various
  SU-LLMs. (b) Area overhead for two PIM designs.}
  \label{fig:challenges_pim}
\end{figure}
\begin{figure}[t]
  \centering
  \includegraphics[width=\linewidth]{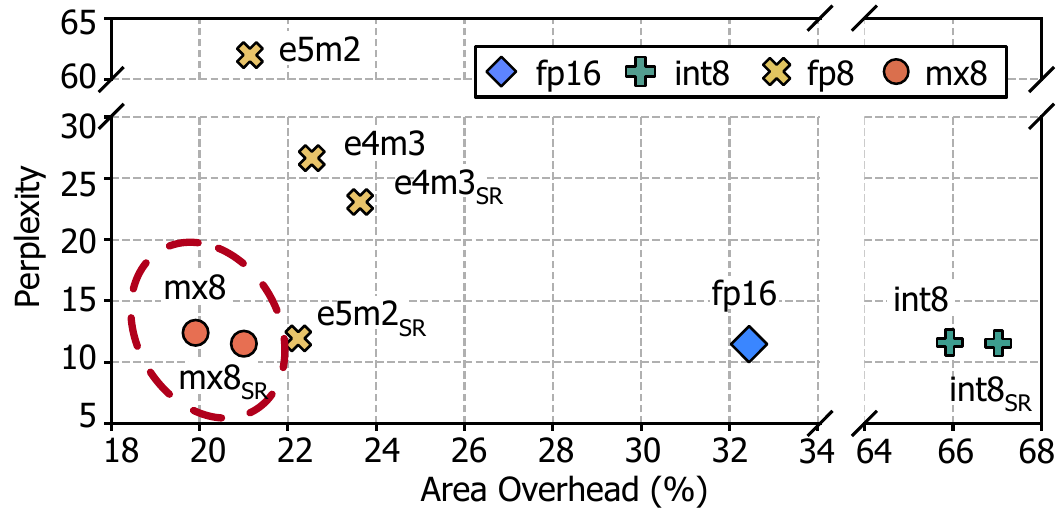}
  \caption{Accuracy-area tradeoff between different low-precision
    formats on Mamba-2 model with WikiText-2. All compute units take
  256-bit-group operands as input.}
  \label{fig:quantization_pareto}
\end{figure}
\begin{figure*}[t]
  \centering
  \includegraphics[width=\linewidth]{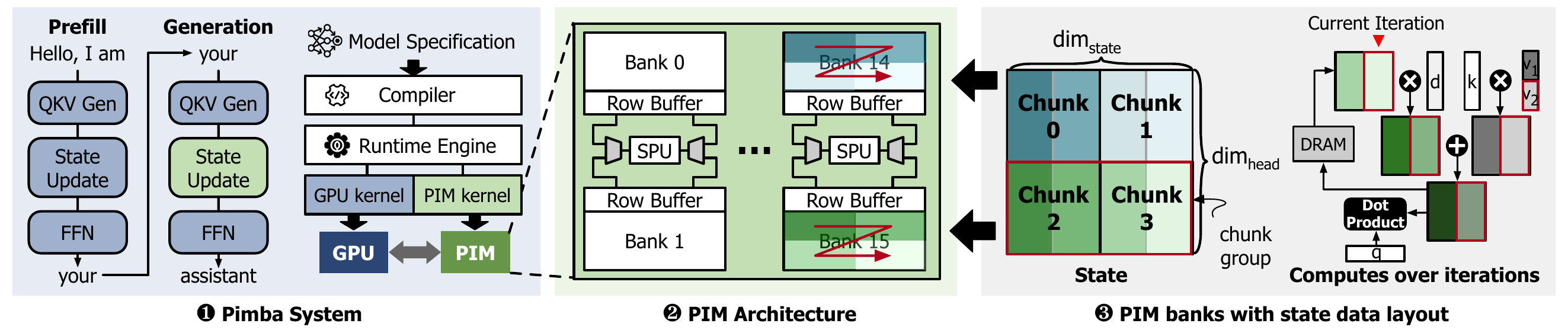}
  \caption{Overview of the proposed \sysname{} system.}
  \vspace{-1ex}
  \label{fig:overview_system}
\end{figure*}

%% file: body/principle.tex
\section{PIM Design Principles}
\label{sec:principles}
Since state update operations are memory-bound, PIM appears to be a
promising solution for acceleration.
While prior works have already explored PIM acceleration for
transformer-based LLM serving~\cite{heo2024neupims,
hyun2024pathfinding, newton, hbm_pim}, we observe that SU-LLMs have
significantly different performance characteristics, necessitating
distinct design decisions.
In this section, we share our empirical insights and the
corresponding principles that govern our accelerator design.

\subsection{Principle 1: Maximizing Hardware Resource Sharing for
Area Efficiency}
\label{subsec:area_throughput}
There are two primary approaches to design PIM for state update
acceleration: (1) time-multiplexed PIM and (2) pipelined PIM.
The time-multiplexed design only implements basic multiplication and
addition units to reduce area overhead, similar to HBM-PIM~\cite{hbm_pim}.
The pipelined design maximizes throughput, by implementing the entire
sequence of operations in a pipelined manner.
\niparagraph{Tradeoff between area and throughput.}
Figure~\ref{fig:challenges_pim} shows the normalized throughput of
state update operations of SU-LLMs on an A100 GPU and the two PIM
designs during the generation phase at batch size 128.
It also presents the respective area overhead of the two PIM designs.
We adopt a per-bank PIM design (each bank equipped with its own
processing unit) for both PIM architectures, matching the channel
bandwidth to that of the A100.
Detailed PIM configurations and simulator/RTL implementation are
provided in Section~\ref{sec:method}.
While the time-multiplexed design offers a modest 2.8$\times$
throughput improvement over the GPU, the pipelined design achieves a
more significant 4.3$\times$ improvement.
Meanwhile, the time-multiplexed design has only 17.8\% area overhead,
whereas the pipelined design incurs a much larger area overhead of
32.4\% (>25\%), posing practical deployment challenges.
Neither design offers both high throughput and low area overhead.
\niparagraph{Achieving both high throughput and area efficiency.}
We argue that even if the number of processing units in the per-bank
pipelined design is halved, the same throughput can still be
maintained, thereby achieving both high throughput and area efficiency.
Even per-bank pipelined designs cannot fully utilize each processing
unit because state updates require both read and write, and row
buffers cannot perform both simultaneously.
During writes, no input is supplied to the processing unit.
With judicious dataflow design, two banks can share a single
processing unit, allowing continuous input from both banks without
throughput loss.
We call this technique \emph{access interleaving} and detail it in
Section~\ref{subsec:access_interleaved_iteration_parallelism}.

\subsection{Principle 2: Achieving Both Accuracy and Efficiency from
Low-Precision Arithmetic}
\label{subsec:area_accuracy}
As discussed in Section~\ref{subsec:challenges_quantization},
quantizing the state is another promising approach to alleviate the
pressure on memory bandwidth during state update operations.
To leverage quantization in PIM, we conduct experiments to evaluate
the tradeoff between area and accuracy aiming to achieve the best of both.
\niparagraph{Tradeoff between area and accuracy.}
Figure~\ref{fig:quantization_pareto} illustrates the area-accuracy
tradeoff space for various low-precision formats, evaluated using the
Mamba-2 on the Wikitext-2~\cite{merity2016pointer} dataset with a
per-bank pipelined PIM design.
Detailed RTL implementation methodology is provided in Section~\ref{sec:method}.
As described in Section~\ref{subsec:challenges_quantization},
\texttt{fp8} formats suffer from a significant increase in
perplexity, while \texttt{int8} and \texttt{MX8} achieve perplexity
levels comparable to \texttt{fp16}.
However, our analysis suggests that \texttt{int8} incurs substantial
area overhead, while \texttt{MX8} is significantly more area-efficient.
The high area overhead of \texttt{int8} stems from the need for
element-wise addition during state updates, which the scaling-based
integer format cannot directly handle.
This necessitates dequantization (multiplying each integer element by
the scaling factor) and re-quantization (normalizing elements based
on the max value), requiring additional arithmetic units and comparison logic.
Conversely, \texttt{MX} simplifies operand alignment for addition by
sharing the exponent bit within a group, enabling direct operations
through simple shifting without dequantization, the details of which
will be discussed in Section~\ref{subsec:microarchitecture}.

Additionally, the results show that stochastic rounding imposes
minimal area overhead.
It requires a Linear Feedback Shift Register (LFSR) for random number
generation and a simple addition unit to add these numbers to the
mantissa, both of which are area-efficient~\cite{9773221}.
We conclude that among the quantized formats, \texttt{MX8} with
stochastic rounding emerges as a Pareto-optimal choice, offering
superior area efficiency while maintaining high accuracy.

%% file: body/architecture.tex
\section{\sysname{}}
\label{sec:arch}

\subsection{Overview}
\label{subsec:system_overview}
Building upon the aforementioned principles, we propose \sysname{}, a
PIM-enabled system designed to accelerate state update and attention
operations while minimizing PIM area overhead.
We first focus on \sysname{} as a state update operation accelerator,
while attention operation acceleration is detailed in
Section~\ref{subsec:attention_support}.
Figure~\ref{fig:overview_system} provides a high-level overview of
\sysname{} system.
\niparagraph{(1) \sysname{} system.}
\sysname{} handles user requests in two phases: prefill and generation.
In the prefill phase, all operations, including state updates, run on
the GPU, as they can be restructured into compute-intensive
forms~\cite{retnet, mamba2}.
In the generation phase, \sysname{} offloads the state update and
attention operations to the PIM, while other operations remain on the GPU.
For the PIM-executed operations, \sysname{} transfers operands to PIM
registers, computes partial sums, and sends the results back to the
GPU for accumulation.
To support heterogeneous execution, \sysname{} includes a software
stack based on a prior work, HBM-PIM~\cite{hbm_pim}.
As in HBM-PIM~\cite{hbm_pim}, \sysname{} device driver first
allocates physically contiguous memory blocks to facilitate
efficient PIM operations.
Custom GPU kernels are implemented for each \sysname{} operation to
issue the necessary PIM commands and compute the addresses of the
memory regions involved in the computations.
To support this, the GPU programming model (e.g., CUDA) is extended
with APIs to issue \sysname{}'s custom DRAM commands.
When GPU kernels are compiled, these APIs are lowered to the
corresponding custom DRAM commands.
These kernels are then registered as custom operations within
high-level frameworks such as PyTorch~\cite{torch2}, allowing users
to invoke \sysname{} functionality seamlessly through familiar APIs.
\niparagraph{(2) PIM architecture.}
\sysname{} aims to accelerate state updates and attention while
minimizing PIM area overhead.
To achieve its objectives, \sysname{} introduces two key architectural
innovations:
(1) \sysname{} employs a novel State-update Processing Unit (SPU) that
is shared between two memory banks
(Section~\ref{subsec:access_interleaved_iteration_parallelism}).
Unlike traditional designs where a processing unit can only access
one bank at a time for either read or write~\cite{hbm_pim},
\sysname{}'s pipelined design allows simultaneous reading from one
bank and writing to the other.
This overlapping of read and write operations across two banks
enables \sysname{} to halve the number of processing units compared
to a per-bank design, while maintaining the same throughput.
(2) Within each SPU, \sysname{} integrates an MX-based State-update
Processing Engine (SPE) to enable area-efficient and
accuracy-preserving computations (Section~\ref{subsec:microarchitecture}).
While prior works primarily focus on MX quantization for dot product
operations~\cite{dacapo}, \sysname{} introduces a microarchitectural
design specifically tailored for MX-based element-wise addition and
multiplication.

\begin{figure}[t]
  \centering
  \includegraphics[width=0.95\linewidth]{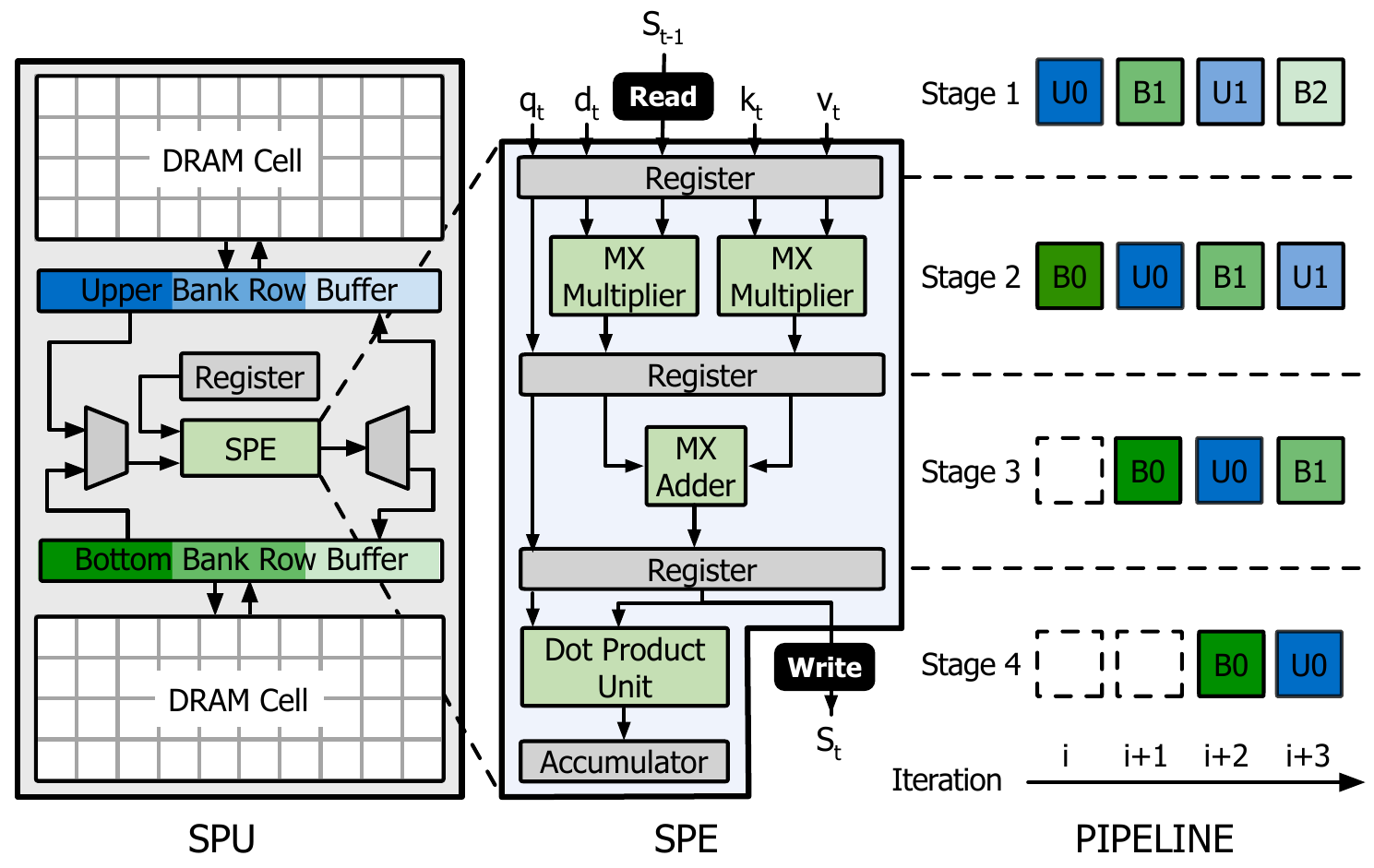}
  \caption{\sysname{} accelerator architecture leveraging access interleaving.}
  \label{fig:Pimba_arch}
\end{figure}
\niparagraph{(3) PIM banks with state data layout.}
We divide each state column along the $dim_{head}$ dimension into
sub-chunks based on the DRAM column size.
Then, we group sub-chunks across the $dim_{state}$ dimension to form
a chunk that aligns with the DRAM row size, enabling efficient
sequential access within each SPU.
To further maximize operand reuse across chunks, we organize the
chunks into chunk groups and assign each group to a DRAM bank.
Chunks within the same group share the operands $d_t$, $q_t$, and
$k_t$, and are placed in consecutive rows of the bank.
This arrangement enables the transfer of shared operands to \sysname{}
once per chunk group, while only the corresponding $v_t$ vector is
transferred per chunk, enhancing data reuse.
Once assigned to banks, \sysname{} processes sub-chunks sequentially in
a pipelined manner by reading consecutive columns within a DRAM row.
In the $i$-th iteration (time unit during which \sysname{} processes
each sub-chunk), \sysname{} utilizes the $i$-th sub-chunk, the shared
$d_t$, $q_t$, $k_t$ vectors, and the $i$-th element of the $v_t$
vector in its computations.

\subsection{Hazard-Free SPU with Access Interleaving}
\label{subsec:access_interleaved_iteration_parallelism}
We propose the access interleaving technique, which maintains the
same throughput as per-bank pipelined designs and reduces area overhead by half.
The core idea, as illustrated in Figure~\ref{fig:Pimba_arch}, is to
have two memory banks share a single SPU, which alternates between
banks in each iteration, enabling continuous data flow by solving the
structural hazard.
At a given iteration, SPU reads a sub-chunk from one bank, referred
to as the upper bank, to initiate a new computation.
Simultaneously, the other bank, referred to as the bottom bank,
performs a write operation to store the result of a computation
initiated several iterations earlier.
This alternating pattern is sustained in the next iteration: the
upper bank now completes a prior computation by writing its result,
while SPU reads a fresh sub-chunk from the bottom bank to begin
another computation.

\begin{figure}[t]
  \centering
  \includegraphics[width=0.9\linewidth]{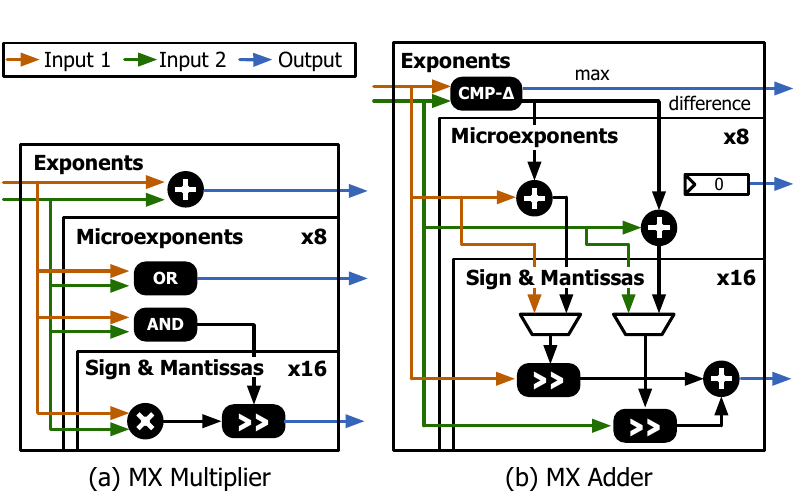}
  \caption{Implementation of MX operations in \sysname{} SPE.}
  \label{fig:mx_operation}
\end{figure}

\niparagraph{Pipeline.}
As depicted in Figure~\ref{fig:Pimba_arch}, \sysname{} employs a
four-stage pipeline:
(1): Fetch the state from DRAM.
(2): Compute state decay and the outer product in parallel.
(3): Sum the results from Stage 2 to update the state.
(4): Perform a dot product between the updated state and $q_t$  while
writing the updated state back to DRAM.
The figure describes the detailed pipelining steps over iteration.
Before iteration $i$, operands are loaded into registers, and chunks
in both upper and bottom banks are activated.
At iteration $i$, sub-chunk $U0$ from the upper bank is read, while
sub-chunk $B0$ from the bottom bank, which is read in a previous
iteration, enters Stage 2.
In iteration $i+1$, sub-chunk $B1$ is read from the bottom bank,
while $U0$ and $B0$ advance to the next stage.
At iteration $i+2$, $U1$ is read from the upper bank, and $B0$ is
written back to the bottom bank.
Notably, at this point, $U1$ is read from the upper bank while $B0$
is written to the bottom bank, avoiding any conflicts due to the use
of separate banks for read and write operations.
Similarly, at iteration $i+3$, $B2$ is read from the bottom bank
while $U0$ is written back to the upper bank.
This interleaving allows continuous, conflict-free processing and
ensures full utilization of compute resources, as SPU receives data
every iteration.

\subsection{Microarchitecture of MX-based SPE}
\label{subsec:microarchitecture}

The MX format, initially designed to accelerate GEMM operations,
requires specific modifications to support element-wise
multiplication and addition.
To this end, we design MX Multiplier and MX Adder, specifically
customized for state update computations.
Each of these computational units operates at three hierarchical
levels: (1) one unit to handle the shared exponent at the group
level, (2) units to manage the microexponents at the sub-group level,
and (3) integer units for each element’s sign and mantissa.

\begin{figure}[t]
  \centering
  \includegraphics[width=\linewidth]{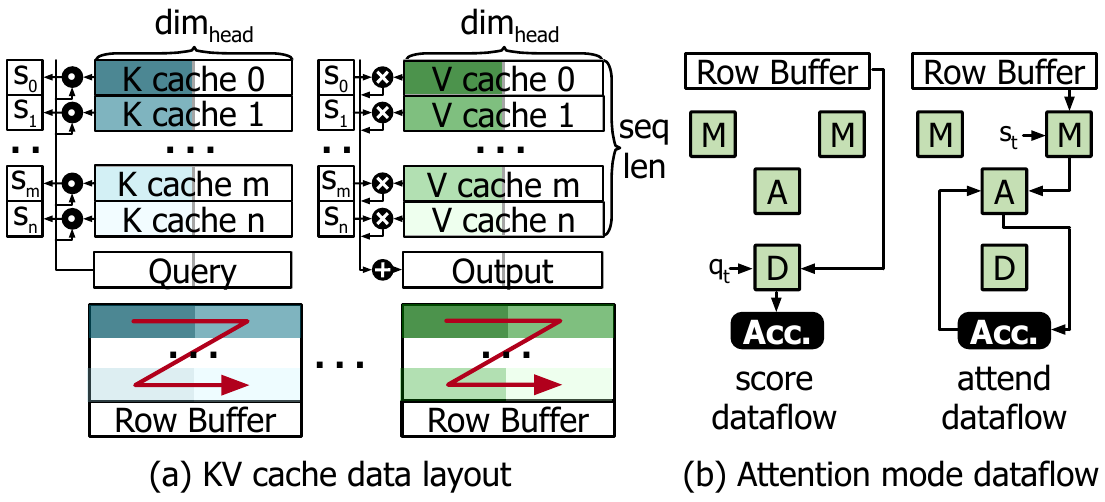}
  \caption{(a) KV cache data layout and (b) attention mode dataflow.
    \textit{M} denotes a multipler, \textit{A} an adder, and \textit{D}
  a dot product unit.}
  \label{fig:attention}
\end{figure}

\niparagraph{MX multiplier.}
Figure~\ref{fig:mx_operation}(a) illustrates how multiplication is
executed with MX.
MX Multiplier adds the exponents of the operand groups to compute the
resulting group exponent.
Similarly, it sums the microexponents within each sub-group.
If the sum of the microexponents exceeds the representable range
(i.e., results in a value of 2, which cannot be represented with 1
bit), it sets the resulting microexponent to 1, and right-shifts the
resulting mantissas of the elements in that sub-group by one, thus
properly adjust the scaling.
The sign and mantissa of each element serve as integers, and are
multiplied using integer multiplication units.
\niparagraph{MX adder.}
Figure~\ref{fig:mx_operation}(b) illustrates how addition is executed with MX.
Addition in MX necessitates an alignment step to ensure operands are
scaled correctly.
MX Adder first aligns the exponents by comparing the two operand
exponents to get their max, which is then used as the resulting exponent.
The group with the smaller exponent adjusts its mantissas by
right-shifting them based on the difference between the exponents.
Additionally, MX Adder further right-shifts the mantissas by their
respective microexponents to ensure proper alignment.
It, then, adds sign and mantissa of each element using integer addition units.
The result of the addition operation always produces a micro-exponent of 0.

\begin{figure}[t]
  \centering
  \includegraphics[width=\linewidth]{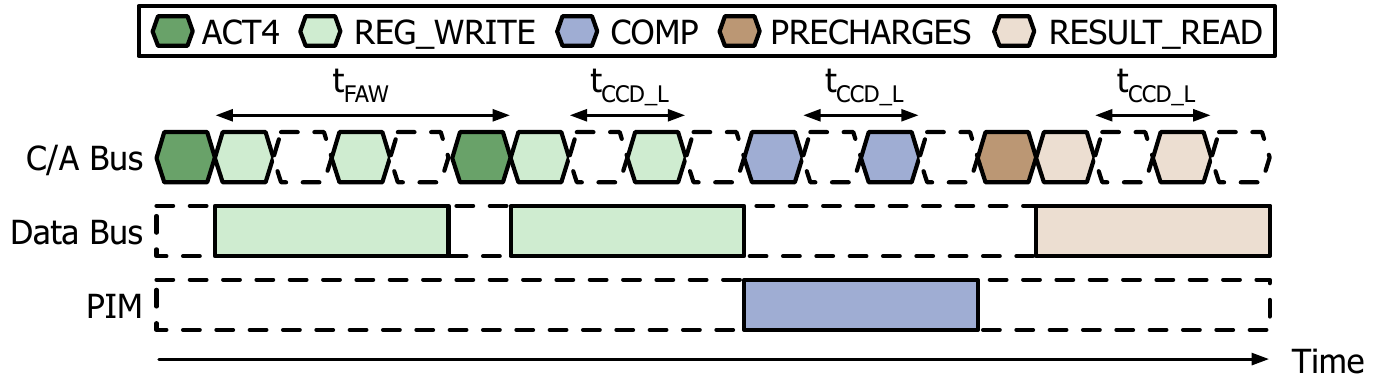}
  \caption{\sysname{} command scheduling. For simplicity, we align C/A
  bus, data bus, and PIM.}
  \label{fig:command_scheduling}
\end{figure}

\subsection{Attention Support in \sysname{}}
\label{subsec:attention_support}
Having described how \sysname{} accelerates state update operations, we
now describe its support for attention operations.
Notably, \sysname{} reuses its existing logic to efficiently execute
attention operations without requiring dedicated hardware extensions.
Figure~\ref{fig:attention}(a) illustrates the layout of the key-value
(KV) cache for attention in \sysname{}.
Similar to the state update data layout, we partition each KV cache
along the $dim_{head}$ dimension into sub-chunks sized to match the
DRAM column width.
These sub-chunks are then grouped into chunks and mapped contiguously
within DRAM rows, preserving spatial locality and reducing access overhead.
The attention computation in \sysname{} proceeds in two phases: score
and attend.
In the score phase, \sysname{} performs dot products between the query
vector and key vectors using its in-pipeline dot product unit as
shown in the score dataflow of Figure~\ref{fig:attention}(b).
The intermediate results are sent to the GPU, where they are
accumulated and passed through a softmax function to produce
normalized attention scores.
In the attend phase, \sysname{} multiplies each attention score with
its corresponding value vector and accumulates the results.
This phase utilizes \sysname{}'s multiplier and adder units, as shown
in the attend dataflow of Figure~\ref{fig:attention}(b).
By leveraging a shared microarchitectural substrate, \sysname{}
achieves efficient support for both state update and attention
operations, demonstrating the versatility of its PIM-based execution model.

\subsection{Memory Interface}
\label{subsec:memory_interface}
In designing \sysname{}, we extend the standard DRAM interface to
enable practical PIM deployment for SU-LLMs.
To achieve this, \sysname{}
(1) adheres to existing DRAM timing constraints to reduce engineering
complexity,
(2) ensures operations perform predefined functions, aligning with
DRAM refresh schemes,
(3) extends the existing DRAM commands to maintain compatibility with
current DRAM, and
(4) employs an all-bank design, as in prior PIMs~\cite{newton,
hbm_pim, gradpim}, to minimize logic overhead.
We propose five custom DRAM commands for state update and attention operations.
\begin{itemize}[leftmargin=*]
  \item{\textbf{ACT4.}}
    The portion of the state and KV cache being processed must first
    be activated to the row buffer.
    Due to power constraints and the $t_{FAW}$ timing window for four
    activations, \sysname{} gangs four activations together, similar to
    previous PIMs~\cite{newton, heo2024neupims}.
  \item{\textbf{REG\_WRITE.}}
    Before operations, \sysname{} receives operands from the host in
    \texttt{MX8} format and stores them in the registers using
    REG\_WRITE command.
    If the host lacks \texttt{MX8} support, Quantization Unit can be
    employed in the host’s memory controller, which can be
    implemented by determining the maximum exponent of incoming
    values and adjusting their mantissas using shift operations,
    enabling a very small area implementation.
  \item{\textbf{COMP.}}
    With states activated and operands loaded, the host initiates
    state update/attention operations across all banks using the COMP command.
    Given that each column read occupies I/O gating, consecutive COMP
    commands must observe the $t_{CCD\_L}$ timing constraint~\cite{gradpim}.
  \item{\textbf{RESULT\_READ.}}
    After computations, the host retrieves results using the
    RESULT\_READ command.
    Since COMP involves both reads and writes for state update
    operations, RESULT\_READ must follow COMP with $t_{RTP}$ and
    $t_{WR}$ intervals.
  \item{\textbf{PRECHARGES.}}
    Through state update operations, the row buffer of each bank
    contains updated state values.
    To store these back into DRAM cells and for the next operations,
    all banks' row buffers are precharged using the PRECHARGES command.
\end{itemize}

\niparagraph{Command scheduling.}
Transferring operands and retrieving results introduce significant
overhead, even with operand reuse.
To reduce this overhead, \sysname{} overlaps data transfer between the
host and \sysname{} during activation and precharge, minimizing this
overhead, as depicted in Figure~\ref{fig:command_scheduling}.
Specifically, the REG\_WRITE command is inserted into the idle time
between ACT4 commands due to the $t_{FAW}$ timing constraint.
Similarly, PRECHARGES takes $t_{RP}$ time to complete, during which
RESULT\_READ is overlapped to reduce the overhead caused by data transmission.

\subsection{Intra- and Inter-\sysname{} Communication}
\label{subsec:interaction}
\niparagraph{Intra-\sysname{} communication.}
\sysname{} attaches the PIM directly to the GPU's off-chip memory,
minimizing communication overhead.
Furthermore, all custom commands executed by \sysname{} operate with
deterministic timing, allowing the GPU to issue multiple PIM
commands in sequence without requiring complex synchronization
logic, as long as timing constraints are satisfied.
Once all PIM operations are completed, the GPU retrieves the
results via a RESULT\_READ command, which delivers the computed
values to the host's execution units through the load queue.
Note that due to data dependencies, GPU and PIM operate in a blocked manner.
For instance, in the attention operation, the GPU remains blocked
until \sysname{} completes all score computations.
Once completed, the results are transferred to the GPU for the
softmax operation, after which \sysname{} resumes to perform the
attend computation.

\input{table/specs}
\begin{figure*}[ht]
  \centering
  \includegraphics[width=0.95\linewidth]{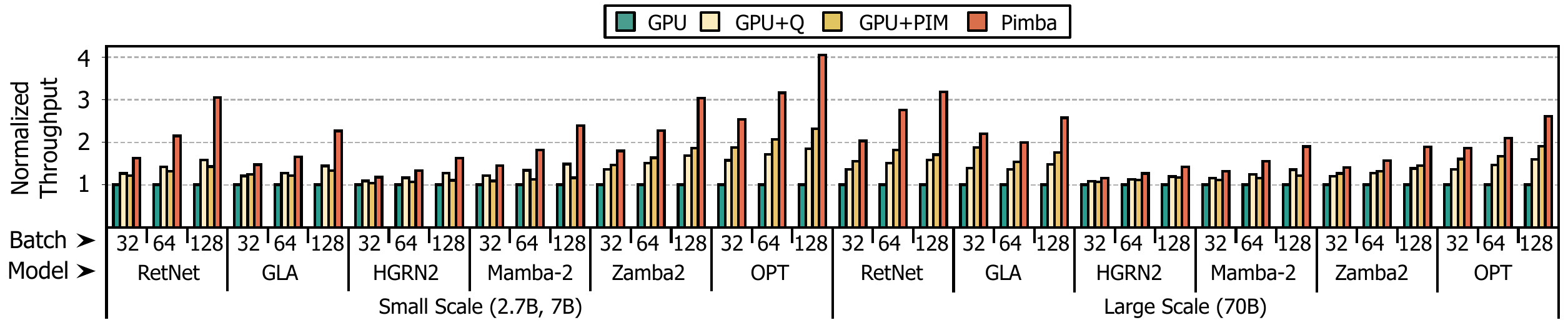}
  \caption{Normalized generation throughput on the baselines and
  \sysname{} on various SU-LLMs and an attention-based LLM.}
  \label{fig:performance_end_to_end}
\end{figure*}
\begin{figure*}[t]
  \centering
  \includegraphics[width=\linewidth]{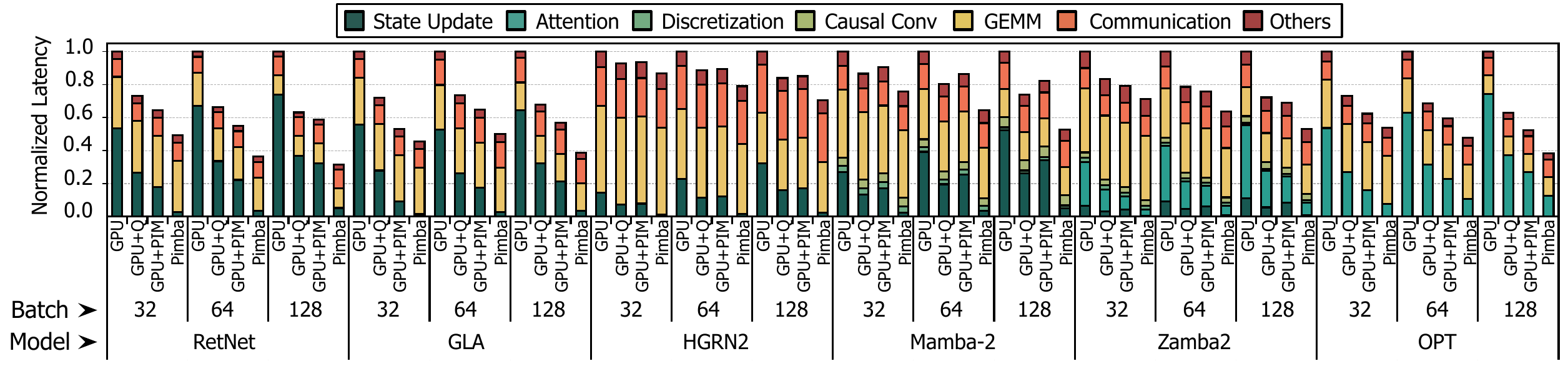}
  \caption{Latency breakdown of large scale SU-LLMs and an
    attention-based LLM at generation phase with (2,048, 2,048)
  input/output sequence lengths.}
  \label{fig:performance_generation}
\end{figure*}
\niparagraph{Inter-\sysname{} communication.}
As LLMs continue to grow in size, a single \sysname{} device cannot
support large-scale models.
\sysname{} addresses this problem by leveraging pipeline and tensor
parallelism, which distributes the model parameters across multiple devices.
To enable such model parallelism, \sysname{} devices exchange
intermediate results via high-bandwidth interconnects such as NVLink.
With pipeline parallelism, the model is partitioned into sequential
blocks, each assigned to a different device, and intermediate
results are forwarded to the next \sysname{} device at block boundaries.
With tensor parallelism, the QKVD projection layers are sharded
along the output channel dimension, and each device computes
partial QKVD vectors.
As state update and attention operate per head, each device
processes only the heads corresponding to its partial vectors.
The outputs are then projected using projection matrices sharded
along the input dimension and aggregated across devices via an
all-reduce operation.
A similar all-reduce is performed after the FFN to complete the
block’s computation.
%

%% file: table/specs.tex
\begingroup
\begin{table}[t]
  \small
  \caption{Specifications of the evaluated HBM.}
  \label{tab:specs}
  \begin{center}
    \begin{tabular}{cc}
      \toprule
      \multicolumn{2}{c}{\textbf{HBM Organization}} \\
      \midrule
      Banks/Bank group & 4 \\
      \midrule
      Bank groups/Pseudo-channel & 4 \\
      \midrule
      Memory Bus Frequency & 1.512GHz \\
      \midrule
      PIM Frequency & {378MHz} \\

      \midrule
      \multicolumn{2}{c}{\textbf{HBM Timing Parameters}} \\
      \midrule
      \multicolumn{2}{c}{tRP = 14, tRAS = 34, tCCD\_S = 2, tCCD\_L =
      4, tWR = 16} \\
      \multicolumn{2}{c}{tRTP\_S = 4, tRTP\_L = 6, tREFI = 3900, tFAW = 30} \\
      \bottomrule
    \end{tabular}
  \end{center}
\end{table}
\endgroup

%% file: body/evaluation.tex
\section{Evaluation}
\label{sec:evaluation}
\subsection{Methodology}
\label{sec:method}

\niparagraph{Models and datasets.}
We evaluate 2.7B parameter SU-LLMs--specifically
RetNet~\cite{retnet}, GLA~\cite{gated_linear_attention},
HGRN2~\cite{hgrn2}, and Mamba-2~\cite{mamba2}--as well as
Zamba2~\cite{zamba2}, a 7B parameter hybrid transformer-Mamba-2 model.
For each model family, we selected the largest publicly available
pretrained architectures.
To provide a baseline with traditional attention-based LLM,
we also evaluate 7B parameter OPT~\cite{zhang2022opt} model.
To quantify the impact of state quantization on model accuracy, we
utilize standard benchmarks widely adopted in LLM evaluations:
WikiText-2~\cite{merity2016pointer}, PIQA~\cite{Bisk2020},
Lambda~\cite{paperno-EtAl:2016:P16-1},
HellaSwag~\cite{zellers2019hellaswag}, ARC-Easy~\cite{allenai:arc},
ARC-Challenge~\cite{allenai:arc}, and Winogrande~\cite{sakaguchi2021winogrande}.
We use perplexity and accuracy as evaluation metrics for WikiText-2
and the others, respectively.
Additionally, to evaluate performance at large scale, we scale the
models to 70B parameters.
Following established practices~\cite{scaling_laws}, we
proportionally scale both the number of layers and hidden dimensions
to reach approximately 70B parameters.
Based on prior findings that increasing state-update head count may
degrade perplexity~\cite{gated_linear_attention}, we retain the
original number of state-update heads and align both $dim_{head}$ and
$dim_{state}$ with the number of heads and hidden dimensions.
\niparagraph{Baselines.}
We evaluate \sysname{} against several baseline systems: an NVIDIA A100
GPU 80GB (\textbf{GPU}), the same GPU configuration but using
\texttt{int8} state quantization matching \sysname{}'s bitwidth
(\textbf{GPU+Q}), and the GPU system with an HBM-PIM~\cite{hbm_pim}
(\textbf{GPU+PIM}).
Both \sysname{} and GPU+PIM systems adopt 40 HBM2E-based PIM memory
modules operating at 1,512MHz, matching the bandwidth of the original
A100 GPU memory.
Given that SPU has a clock cycle of $t_{CCD\_L}$ (4 memory
bus cycles), its frequency is 378MHz, which is consistent with prior
work~\cite{hbm_pim}.
The HBM-PIM is designed using a time-multiplexed design that places a
\texttt{fp16} processing unit spanning two banks without the access
interleaving technique of \sysname{}.
This results in an area overhead comparable to that of \sysname{} PIM.
For small scale models, all evaluated systems utilize a single GPU,
as these models comfortably fit within one GPU's memory capacity.
For large scale models, all systems employ eight GPUs interconnected
through a high-bandwidth network analogous to the NVIDIA DGX A100 system.
We use NVLink3 as the interconnect for GPU-to-GPU
communication, providing a bandwidth of 600GB/s, and the models are
partitioned using tensor parallelism.
\niparagraph{Cycle-accurate simulator.}
We develop an in-house cycle-accurate simulator based on
Ramulator2~\cite{luo2023ramulator2} to evaluate the performance and
energy efficiency of the PIM subsystem, incorporating existing DRAM
timing constraints and refresh schemes.
We model other system components including GPUs and NVLink by
extending an open source simulator~\cite{park2024attacc}.
We refer to the activation and read energy of HBM from the previous
work~\cite{fine_grained_dram}.
Detailed HBM configurations are presented in Table~\ref{tab:specs}.
\niparagraph{Area and power.}
We synthesize \sysname{} accelerator using Synopsys Design Compiler
with the FreePDK 45nm technology node~\cite{freepdk}, scaling area
and power values to 10nm using the DeepScaleTool~\cite{9401196}.
The same procedure is applied to the HBM-PIM, with components sourced
from the Synopsys DesignWare libraries.
Scaling follows methods from prior PIM works, considering that memory
processes are 10$\times$ less dense than logic processes of the same
feature size~\cite{park2024attacc}.
SRAM-based buffers are modeled using CACTI7~\cite{cacti7} at 22nm,
then scaled to 10nm.

\input{table/accuracy}

\subsection{Results}
\label{subsec:result}
\niparagraph{Throughput.}
Figure~\ref{fig:performance_end_to_end} reports the normalized
token-generation throughput across various SU-LLMs and an
attention-based LLM using (2,048, 2,048) input/output sequence lengths.
The GPU+Q system achieves an average of 1.4$\times$ higher
throughput over the GPU baseline due to halving the state size.
Interestingly, GPU+PIM also achieves 1.4$\times$
throughput improvement over GPU, matching or occasionally
underperforming compared to GPU+Q despite leveraging PIM.
This occurs because HBM-PIM employs a time-multiplexed design without
\sysname{}’s access interleaving, causing state update operations to
span multiple cycles.
Consequently, even with PIM, internal bandwidth is underutilized,
yielding sub-optimal performance.
Furthermore, GPU+PIM processes twice as much data compared to GPU+Q,
as it uses \texttt{fp16} states rather than 8-bit formats.
In contrast, \sysname{} consistently outperforms GPU and GPU+PIM,
providing average throughput gains of 1.9$\times$ and
1.4$\times$ (up to 4.1$\times$ and 2.1$\times$), respectively.
This superior performance results from efficiently leveraging
internal bandwidth through pipelining and access interleaving,
alongside the advantages of low-precision state representation.

\begin{figure}[t]
  \centering
  \includegraphics[width=\linewidth]{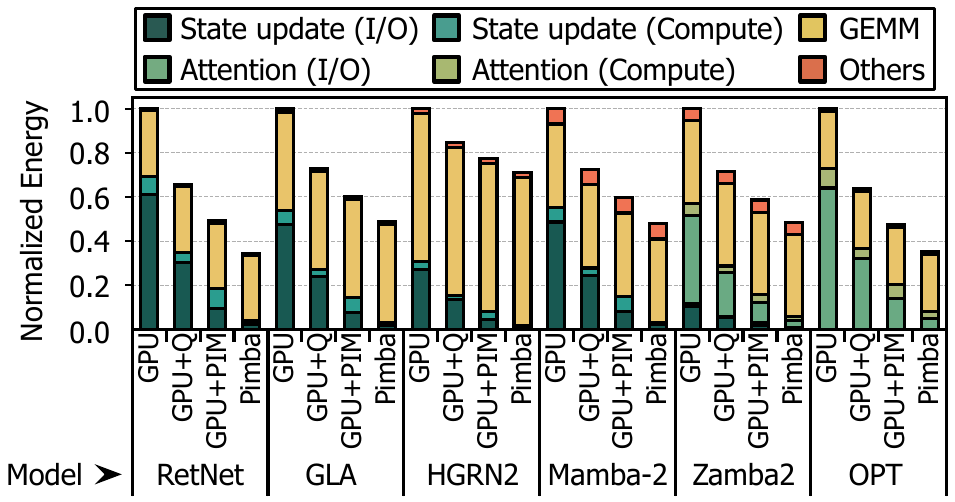}
  \caption{Normalized energy consumption of large scale SU-LLMs
    and an attention-based LLM at generation phase with
  batch size 128.}
  \label{fig:eval_power}
\end{figure}

\niparagraph{Latency breakdown.}
To identify the main contributors to throughput improvements,
Figure~\ref{fig:performance_generation} presents normalized latency
for large scale SU-LLMs and an attention-based LLM during
the generation phase.
As shown, \sysname{} reduces state update operation latency by
14.6$\times$ and 6.9$\times$ compared to GPU and GPU+PIM,
respectively, highlighting its efficient handling of state updates.
Additionally, the results show greater overall latency reduction for
larger batch sizes and models dominated by state update operations.
For instance, HGRN2 with a batch size of 32 spends only 14\% of its
latency on state updates, resulting in a modest 1.2$\times$ latency reduction.
In contrast, RetNet with a batch size of 128 allocates 74\% of its
latency to state updates, achieving a substantial 3.2$\times$ latency reduction.
Moreover, \sysname{} achieves latency reductions for attention
operations by 6.3$\times$ and 2.1$\times$ compared to GPU and
GPU+PIM, respectively, effectively accelerating
attention-dominated hybrid models such as Zamba2 and the
attention-based OPT.
Notably, the latency reduction for attention operations in \sysname{}
is smaller compared to state updates.
This is due to attention operations primarily consisting of GEMV
computations without requiring write operations, thus limiting the
benefits of \sysname{}’s pipelined design and access interleaving.
Nevertheless, by employing the \texttt{MX8} format, \sysname{} achieves
a 2.1$\times$ attention latency reduction over GPU+PIM,
demonstrating robust performance in hybrid models as well.

\input{table/area}

\niparagraph{Energy consumption.}
Figure~\ref{fig:eval_power} presents the normalized energy
consumption breakdown of large scale SU-LLMs and an
attention-based LLM with a batch size of 128 during the generation phase.
The results show that \sysname{} achieves an average of
2.2$\times$ lower energy consumption compared to GPU,
demonstrating its cost-effectiveness for SU-LLM serving.
These improvements stem from reducing the amount of state and KV
cache transfer between the GPU and memory while performing state
update operations within the PIM itself.
Additionally, \sysname{} exhibits an average of 1.3$\times$ lower
energy consumption compared to GPU+PIM, attributed to \sysname{}’s use
of the \texttt{MX8} which significantly reduces transfer between the
GPU as well as computation time.

\niparagraph{Accuracy.}
Table~\ref{tab:accuracy} summarizes the accuracy results comparing
\sysname{} with the GPU baseline for various small scale SU-LLMs.
Experimental results indicate that despite the use of \texttt{MX8}
quantization, \sysname{} achieves comparable accuracy to the GPU baseline.
This is because the use of \texttt{MX8}, which has a high mantissa
bit-width, combined with stochastic rounding, significantly mitigates
swamping effects during continuous state updates.
In conclusion, given the hardware efficiency of \texttt{MX8}, the
results affirm that \texttt{MX8} is an attractive choice to reduce
area overhead for implementing SPE, while offering a negligible
impact on the LLM accuracy.
\niparagraph{Area.}
To assess the practicality of \sysname{}, we measure the area and power
consumption of \sysname{} and HBM-PIM.
Table~\ref{tab:area} presents the results.
For a fair comparison, HBM-PIM is optimized for state update
computations by retaining only the essential components while
reducing or removing others, such as shrinking internal registers and
eliminating specific control logic.
As indicated in the table, \sysname{} adheres to PIM area constraints,
with an area overhead of 13.4\%, well below the 25\% maximum logic
ratio recommended by prior work~\cite{newton}.
Although \sysname{} incurs an area overhead approximately 1.5\% larger
than HBM-PIM, this increase is justified by delivering up to a
2.1$\times$ throughput improvement over HBM-PIM.
\begin{figure}[t]
  \centering
  \includegraphics[width=0.9\linewidth]{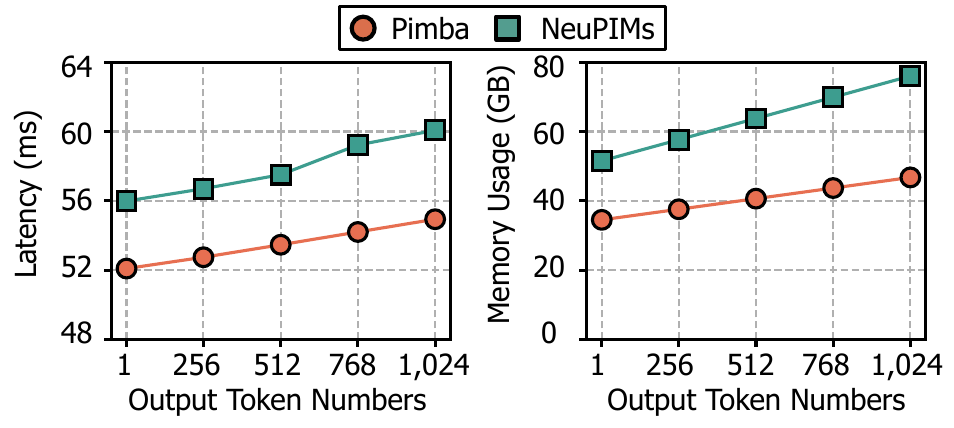}
  \caption{Latency and memory usage of NeuPIMs and \sysname{} as the
  number of generated output tokens increases with batch size 128.}
  \label{fig:neupims_versus}
\end{figure}
\begin{figure}[t]
  \centering
  \includegraphics[width=\linewidth]{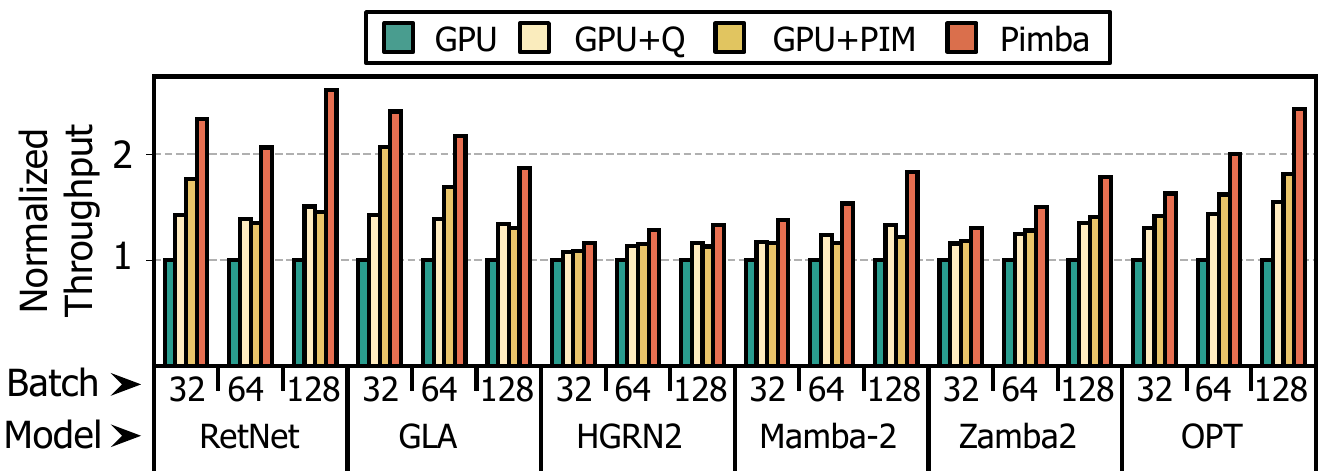}
  \caption{Normalized generation throughput on the
  baselines and \sysname{} with NVIDIA H100 GPU settings.}
  \label{fig:performance_h100}
\end{figure}
\niparagraph{Comparison with existing PIM-enabled system.}
We compare the latency and memory usage of \sysname{} with
NeuPIMs~\cite{heo2024neupims}, a per-bank PIM-based acceleration
solution tailored for attention operations.
We use the Zamba2 70B model with batch size 128 and (1,024, 1,024)
input/output lengths.
Both systems use eight A100 GPUs with tensor parallelism.
To ensure fairness, we match the timing parameters and the number of
HBM stacks across both systems.
Figure~\ref{fig:neupims_versus} demonstrates that \sysname{}
consistently achieves lower latency compared to NeuPIMs.
This improvement arises primarily because \sysname{} efficiently
offloads state update operations onto the PIM, a critical capability
lacking in NeuPIMs.
Moreover, \sysname{} exhibits latency scaling similar to NeuPIMs as the
number of output tokens increases, despite utilizing half the number
of processing units, not employing dual row buffers, and foregoing
sub-batch interleaving techniques present in NeuPIMs.
This is enabled through optimized command scheduling and the adoption
of low precision representations for the KV cache.
Additionally, \sysname{} consistently maintains lower memory usage
compared to NeuPIMs, further underscoring the effectiveness of
employing low precision representations for both state data and KV cache.
\niparagraph{General adoption of \sysname{}.}
While our initial \sysname{} is based on the NVIDIA A100 GPU, it can
be integrated with any GPU.
To demonstrate this general applicability,
Figure~\ref{fig:performance_h100} presents the normalized
throughput of large scale LLMs using both baseline systems and
\sysname{} under the NVIDIA H100 GPU configuration.
All systems adopt 40 HBM3-based PIM modules operating at a memory
bus frequency of 2.626GHz and an SPU frequency of 657MHz, matching
the memory bandwidth of the H100.
They also use NVLink4 for GPU-to-GPU communication, which provides
900GB/s of bandwidth.
Under this setting, \sysname{} exhibits a similar acceleration trend
to that observed on the A100, consistently outperforming both the
GPU and GPU+PIM baselines by 1.8$\times$ and 1.3$\times$ on average.
These results demonstrate both the applicability and effectiveness
of \sysname{}, regardless of the underlying GPU platform.

%% file: table/accuracy.tex
\begingroup
\begin{table*}[t]
  \caption{Accuracy evaluation for various small scale SU-LLMs and an
  attention-based LLM}
  \label{tab:accuracy}
  \begin{center}
    \begin{tabular}{llcccccccc}
      \toprule[1pt]
      \multirow{2}{*}{\vspace{-1.5ex}Model} &
      \multirow{2}{*}{\vspace{-1.5ex}Method} &
      Perplexity $\downarrow$ &
      \multicolumn{7}{c}{Accuracy $\uparrow$ (\%)} \\
      \cmidrule(lr){3-3} \cmidrule(lr){4-10}
      & & WikiText-2 & Piqa & Lambada & HellaSwag & ARC-E & ARC-C &
      WinoGrande & Geomean \\

      \midrule[1pt]
      \multirow{2}{*}{RetNet} & GPU & \textbf{15.83} & \textbf{72.3}
      & \textbf{44.0} & \textbf{42.0} & 59.5 & 25.5 &
      53.1 & 47.0 \\ & \sysname{} & 15.95 &
      \textbf{72.3} & 43.7 & 41.9 & \textbf{59.7} & \textbf{25.8} &
      \textbf{53.7} & \textbf{47.1}~\textcolor{black}{(+0.1)} \\

      \midrule
      \multirow{2}{*}{GLA} & GPU & \textbf{15.54} & \textbf{71.6} &
      \textbf{43.8} & 41.8 & \textbf{59.1} & \textbf{26.7} &
      \textbf{55.4} & \textbf{47.5} \\ &
      \sysname{} & 15.57 & 71.5 & 43.4 & \textbf{41.9} & \textbf{59.1}
      & \textbf{26.7} & 55.2 &
      47.4~\textcolor{black}{(--0.1)}\\

      \midrule
      \multirow{2}{*}{HGRN2} & GPU & \textbf{14.48} &
      \textbf{73.1} & \textbf{48.5} & 44.6 & \textbf{60.7} & \textbf{25.3} &
      54.7 & \textbf{48.6} \\
      & \sysname{} & 15.09 & 73.0 & \textbf{48.5} &
      \textbf{44.8} & 60.3 & 25.2 & \textbf{55.0} &
      \textbf{48.6}~\textcolor{black}{(+0.0)}\\

      \midrule
      \multirow{2}{*}{Mamba-2} & GPU & \textbf{11.46} &
      \textbf{76.4} & \textbf{59.6} & \textbf{49.6} & \textbf{69.4} &
      \textbf{33.2} & \textbf{64.0} & \textbf{56.7} \\
      & \sysname{} & 11.51 & 76.3 & 59.4 & \textbf{49.6} & 69.3 &
      \textbf{33.2} & 63.9 & 56.6~\textcolor{black}{(--0.1)} \\

      \midrule
      \multirow{2}{*}{Zamba2} & GPU & \textbf{5.94} &
      \textbf{78.9} & \textbf{64.9} & \textbf{63.8} & \textbf{78.9} &
      \textbf{53.8} & \textbf{77.7} & \textbf{69.0} \\
      & \sysname{} & 5.96 & 78.6 & 64.4 & 63.7 & 78.4
      & \textbf{53.8} & 77.1 & 68.7~\textcolor{black}{(--0.3)}\\

      \midrule
      \multirow{2}{*}[-2pt]{OPT} & GPU & \textbf{12.29} & \textbf{76.2} &
      \textbf{63.3} & \textbf{50.5} & \textbf{65.6} & \textbf{30.6} & 65.1 &
      \textbf{56.3} \\ & \sysname{} & \textbf{12.29} & 76.1 & 63.2 &
      \textbf{50.5} & \textbf{65.6} & 30.3 & \textbf{65.2} &
      56.2~\textcolor{black}{(--0.1)} \\

      \bottomrule[1pt]
    \end{tabular}
  \end{center}
\end{table*}
\endgroup

%% file: table/area.tex
\begingroup
\begin{table}[t]
  \small
  \caption{Area and power comparison.}
  \label{tab:area}
  \begin{center}
    \begin{tabular}{ccc}
      \toprule[1pt]
      \textbf{Parameters} & \textbf{\sysname{}} & \textbf{HBM-PIM} \\

      \midrule[1pt]
      Compute area ($mm^2$) & 0.053 & 0.042 \\
      \midrule
      Buffer area ($mm^2$) & 0.039 & 0.039 \\
      \midrule
      Total area ($mm^2$) & 0.092 & 0.081 \\
      \midrule
      Area overhead (\%) & 13.4 & 11.8 \\
      \midrule
      Compute power dissipation (mW) & 8.2908 & 6.028 \\
      \bottomrule[1pt]
    \end{tabular}
  \end{center}
\end{table}
\endgroup

%% file: body/related_works.tex
\section{Related Work}
\niparagraph{Post-transformer accelerators.}
As post-transformer LLMs gain traction, the architecture community
has initiated efforts to accelerate these models.
Yoon et al.~\cite{yoon24analysish3} conducts a detailed
characterization of the Hungry Hungry Hippos (H3) model, a variant of
SSM, and highlights the generation phase as a major bottleneck.
VGA~\cite{vga} identifies inefficiencies in Fast Fourier Transform
(FFT) operations on GPUs in the H3 model and proposes an FFT-based
convolution architecture to address these inefficiencies.
MARCA~\cite{marca} focuses on accelerating the Mamba model, the
predecessor of the Mamba-2 model, by introducing architectural
optimizations for element-wise operations and nonlinear computations
used in the model.
However their scope is limited to SSM-based models, and even within
SSMs, they target specific instances such as the H3 or Mamba models.
In contrast, our work represents the first comprehensive analysis of
a wide range of post-transformer models.
We rigorously examine the computational patterns shared across many
post-transformers and propose PIM-based acceleration that is broadly applicable.

\niparagraph{PIM-based LLM serving systems.}
Recently, there has been significant research on PIM-based LLM
serving
systems~\cite{alsop2024inclusivepimhardwaresoftwarecodesignbroad,
  hyun2024pathfinding, heo2024neupims, ianus, psyncpim, llmservingsim,
newton, hbm_pim}.
However, these systems predominantly target transformer-based LLMs,
focusing on GEMV operations.
Furthermore, they provide a limited exploration of integrating
quantization techniques into PIM architectures.
In contrast, \sysname{} targets both state update and attention
operations for post-transformer LLM acceleration.
To address the increased area overhead associated with this
versatility, \sysname{} employs resource sharing and quantization techniques.

\niparagraph{Quantization.}
Quantization techniques have garnered significant attention in both
the algorithm and system communities as an effective solution for
accelerating LLMs~\cite{xiao2023smoothquant, hooper2024kvquant, monkey, atom}.
Recently, research on quantization for post-transformer models has
been actively pursued.
Quamba~\cite{quamba} and MambaQuant~\cite{mambaquant} propose a
technique for quantizing both activations and weights of the Mamba
model to 8-bit.
However, these works focus exclusively on activation and weight
quantization and do not address state quantization, making them
orthogonal to our research.
Q-Mamba~\cite{qmamba} proposes Decoupled Scale Quantization, which
employs an \texttt{int8} format with decoupled scales for state
quantization in the Mamba model.
Although this approach is effective, it necessitates additional
arithmetic operations and logic to manage scaling factors, thus
limiting its practicality for PIM architectures.
On the other hand, our work identifies that MX
format~\cite{rouhani2023shared} achieves Pareto optimality in terms
of the area-accuracy trade-off for PIM environments, and we
incorporate this format into our design.
%

%% file: body/discussion.tex
\section{Discussion}
\label{sec:discussion}
\niparagraph{Improving utilization.}
The operations in SU-LLM must be executed sequentially, leading to
inherent data dependencies.
To ensure correct ordering under these dependencies, GPU and PIM in
\sysname{} alternate their execution in a blocked manner.
This leads to underutilization of both GPU and PIM resources,
degrading overall performance.
Recent work, NeuPIMs~\cite{heo2024neupims}, proposes dual row
buffers and sub-batch interleaving to address this issue.
These techniques enable concurrent read/write operations and PIM
command execution, and allow overlapping GPU and PIM execution
across two sub-batches, thereby improving utilization.
Note that the NeuPIMs techniques are orthogonal to our approach, as
they focus primarily on system-level scheduling and architectural
modifications to the row buffer design.
Integrating such techniques into \sysname{} could effectively
eliminate execution bubbles between GPU and PIM, thereby further
improving overall resource utilization.
%

%% file: body/conclusion.tex
\section{Conclusion}

This paper presents \sysname{}, a Processing-in-Memory (PIM)
accelerator designed to efficiently serve both transformer and
post-transformer LLMs under the increasing demands of long-context,
high-throughput inference.
Through detailed workload characterization, we identify that
\emph{state update}--a central operation in post-transformer
models--shares similar memory bandwidth bottlenecks with attention in
transformer-based models, motivating a unified acceleration approach.
\sysname{} addresses these challenges by combining in-memory
computation with quantized execution, co-designing its architecture
around two key principles: (1) maximizing hardware resource sharing
to reduce area cost, and (2) selecting Pareto-optimal quantization
formats for efficient and accurate execution.
Our evaluation shows that fine-grained access interleaving and
MX-based state update engines enable \sysname{} to deliver significant
improvements in throughput and area efficiency.
These results suggest \sysname{}'s potential as a practical and
generalizable solution for next-generation LLM serving infrastructure.

%% file: body/ack.tex
\begin{acks}

  %
  We thank the anonymous reviewers for their comments and feedback.
  This work was supported by the Institute of Information \&
  Communications Technology Planning \& Evaluation (IITP)
  (No.RS-2024-00396013, No.2022-0-01037), IITP under the Graduate School of
  Artificial Intelligence Semiconductor (IITP-2025-RS-2023-00256472),
  grant funded by the Korea government (MSIT).
  This work was also supported by Electronics and Telecommunications
  Research Institute (ETRI) grant funded by ICT R\&D program of
  MSIT/IITP (No.RS-2025-02305453, Development of distributed
    inference and model optimization technology for heterogeneous Al
  semiconductors).
  The EDA tool was supported by the IC Design Education Center (IDEC), Korea.
\end{acks}

%% file: body/ae.tex
\appendix

\section{Artifact Appendix}

\subsection{Abstract}
In this work, we present \sysname{}, a processing-in-memory
acceleration solution for post-transformer large language models.
Our artifact includes a full-system simulator for \sysname{} as well as
accuracy evaluation code.
To ensure reproducibility and streamline the experimental workflow,
we made two key engineering efforts:
(1) The entire codebase relies solely on the modern project manager
\texttt{uv}, allowing evaluators to install exactly the same versions
of dependencies and build tools with a one-line command; and
(2) we provide only two scripts--one to run all experiments and
another to generate all figures from the results--to reduce the burden
on evaluators.
These two engineering efforts collectively enable a streamlined
experimental workflow and maximize reproducibility.
Beyond the artifact evaluation, we also document the challenges we
encountered while preparing
the code and a breif API reference to assist those who wish to extend
our codebase.
We believe these resources will help those preparing artifact
evaluations and aid future research efforts.

\subsection{Artifact check-list}

\begin{itemize}[leftmargin=*]
  \item \textbf{Program:} uv
  \item \textbf{Compilation:} gcc
  \item \textbf{Model:} Our code automatically downloads all required
    model weights from Hugging Face: RetNet, GLA, HGRN2, Mamba2,
    Zamba2, OPT, LLaMA
  \item \textbf{Data sets:} Our code automatically downloads all required
    datasets from Hugging Face: Wikitext2, Piqa, Lambada,
    Hellaswag, Arc-easy, Arc-challenge, Winogrande
  \item \textbf{Run-time environment:} CUDA, x86\_64
  \item \textbf{Hardware:} An NVIDIA GPU based on the Ampere
    architecture or newer, equipped with at least 24GB of memory
  \item \textbf{Metrics:} Perplexity, Accuracy, Throughput
  \item \textbf{Output:} PDF files for the figures, a CSV file for the table
  \item \textbf{How much disk space required (approximately)?:} The
    installation occupies about 100GB in the user's "\$HOME" directory,
    primarily due to models and datasets stored under
    "\$HOME/.cache/huggingface".
  \item \textbf{How much time is needed to prepare workflow
    (approximately)?:} 10 minutes
  \item \textbf{How much time is needed to complete experiments
    (approximately)?:} 15 hours
  \item \textbf{Publicly available?:}
    \bluetext{\url{https://github.com/casys-kaist/pimba}}
  \item \textbf{Archived (provide DOI)?:}
    \bluetext{\url{https://doi.org/10.5281/zenodo.16946084}}
\end{itemize}

\subsection{Description}
\niparagraph{How to access.}
Clone our public repository from github:
\begin{verbatim}
  $ git clone https://github.com/casys-kaist/pimba
\end{verbatim}
\niparagraph{Hardware dependencies.}
Evaluators require an NVIDIA GPU based on
the Ampere architecture or newer, with at least 24GB of memory
(e.g., RTX 3090, RTX 4090, RTX 5090, A6000, A100, etc.).
\niparagraph{Software dependencies.}
Evaluators only need \texttt{gcc} and \texttt{uv}; all other required
python packages and build tools are installed through \texttt{uv}.
\texttt{uv} itself can be installed using the system package manager
or via a one-line command provided in the official documentation~\cite{uv}.
\niparagraph{Models and data sets.}
During the experiments, our code automatically downloads all required
models and datasets.

\subsection{Installation}
With the following commands, \texttt{uv} automatically downloads the required
dependencies and build tools (e.g., cmake, ninja) in exactly the same
versions we used and compiles our project.
\begin{verbatim}
  $ uv sync
  $ uv run cmake --preset release
  $ uv run cmake --build build
\end{verbatim}

\subsection{Experiment workflow}
Evaluators can run all experiments with a single command:
\begin{verbatim}
  $ uv run python scripts/run.py
\end{verbatim}
This command generates two files, "accuracy\_result.yaml" and
"performance\_result.yaml", under the "res/" directory, which are
then used to reproduce the figures and the table.

\subsection{Evaluation and expected results}
Using the result files, evaluators can simply reproduce the figures
and the table in the paper by executing the following command:
\begin{verbatim}
  $ uv run python scripts/draw.py
\end{verbatim}
This process generates PDF files for the figures and a CSV file for
the table in the "summary/" directory.
Note that the PDF files do not include axis labels or legends, as we
add them using external drawing tools.
However, all scales and colors precisely match those in the paper,
ensuring no issues arise when
comparing values.
We also observed that accuracy may slightly vary with the GPU
platform and CUDA version used for evaluation, especially in
Table~\ref{tab:accuracy}.
However, the key trend we aim to convey in this paper remains
consistent: quantization with \texttt{MX8} has minimal impact on accuracy.